\documentclass[a4paper,11pt]{article}
\pdfoutput=1
\usepackage{jheppub}

\usepackage{textcase}
\usepackage{graphicx,epsfig,psfrag,amssymb,hyperref}
\usepackage{multirow}
\usepackage{color,graphicx,epsfig,psfrag,amsmath,empheq}
\usepackage{bm}
\usepackage{mathrsfs,amsfonts,soul,color}
\usepackage[caption=false]{subfig}
\usepackage{hepunits}

\usepackage{placeins}

\newcommand{\tev}{\ensuremath{\mathrm{\: Te\kern -0.1em V}}\xspace}
\newcommand{\gev}{\ensuremath{\mathrm{\: Ge\kern -0.1em V}}\xspace}
\newcommand{\mev}{\ensuremath{\mathrm{\: Me\kern -0.1em V}}\xspace}

\def\aprime{\ensuremath{A^{\prime}}\xspace}
\def\ma{\ensuremath{m_{A^{\prime}}}\xspace}
\def\maeps {\ensuremath{[\ma,\varepsilon^2]}\xspace}

\def\BL {\ensuremath{B\!-\!L}\xspace}
\def\pphob {\ensuremath{\text{\st{$p$}}}\xspace}

\newcommand{\cL}{\mathcal{L}}
\newcommand{\cF}{\mathcal{F}}
\newcommand{\cB}{\mathcal{B}}
\newcommand{\cR}{\mathcal{R}}
\newcommand{\cA}{\mathcal{A}}
\newcommand{\cK}{\mathcal{K}}
\newcommand{\cO}{\mathcal{O}}

\def\git {\href{https://gitlab.com/philten/darkcast}{\tt https://gitlab.com/philten/darkcast}}

\begin{document}

\preprint{MIT-CTP/4976, CERN-TH-2017-282}

\title{Serendipity in dark photon searches}

\author{Philip Ilten,$^a$}
\author{Yotam Soreq,$^b$}
\author{Mike Williams,$^c$}
\author{and Wei Xue$^d$}

\affiliation{\phantom{ }\hspace{-0.12in}$^a$School of Physics and Astronomy, University of Birmingham, Birmingham, B152 2TT, UK\\
\phantom{ }\hspace{-0.16in}$^b$Center for Theoretical Physics, Massachusetts Institute of Technology, \!Cambridge, \!MA 02139, \!USA \\
\phantom{ }\hspace{-0.16in}$^c$Laboratory for Nuclear Science, Massachusetts Institute of Technology, \!Cambridge, \!MA 02139, \!USA \\
\phantom{ }\hspace{-0.16in}$^d$Theoretical Physics Department, CERN, CH-1211 Geneva 23, Switzerland
}

\emailAdd{philten@cern.ch}
\emailAdd{soreqy@mit.edu}
\emailAdd{mwill@mit.edu}
\emailAdd{wei.xue@cern.ch}

\abstract{
Searches for dark photons provide serendipitous discovery potential for other types of vector particles.
We develop a framework for recasting dark photon searches to obtain constraints on more general theories, which includes a data-driven method for determining hadronic decay rates.
We demonstrate our approach by deriving constraints on a vector that couples to the \BL current, a leptophobic $B$ boson that couples directly to baryon number and to leptons via $B$--$\gamma$ kinetic mixing, and on a vector that mediates a protophobic force.
Our approach can easily be generalized to any massive gauge boson with vector couplings to the Standard Model fermions, and software to perform any such recasting is provided at \git.
}

\maketitle

\section{Introduction}
\label{sec:intro}

Substantial effort has been dedicated in recent years~\cite{Essig:2013lka,Alexander:2016aln,Battaglieri:2017aum} to searching for a massive {\em dark photon}, \aprime, whose small coupling to the electromagnetic~(EM) current arises due to kinetic mixing between the Standard Model~(SM) hypercharge and \aprime field strength tensors~\cite{Okun:1982xi,Galison:1983pa,Holdom:1985ag,Pospelov:2007mp,ArkaniHamed:2008qn,Bjorken:2009mm}.
This mixing provides a potential portal through which dark photons may be produced in the lab, and also via which they can decay into visible SM final states---though decays into invisible dark-sector final states are expected to be dominant if kinematically allowed.

The minimal \aprime model has 3 unknown parameters: the mass of the dark photon, \ma; the kinetic-mixing strength, $\varepsilon^2$; and the dark photon decay branching fraction into invisible dark-sector final states, which is typically assumed to be either 0 or $\approx\!1$.
Constraints have been placed on visible \aprime decays by previous beam-dump~\cite{Bergsma:1985is,Konaka:1986cb,Riordan:1987aw,Bjorken:1988as,Bross:1989mp,Davier:1989wz,Athanassopoulos:1997er,Astier:2001ck,Bjorken:2009mm,Essig:2010gu,Williams:2011qb,Blumlein:2011mv,Gninenko:2012eq,Blumlein:2013cua,Banerjee:2018vgk},
fixed-target~\cite{Abrahamyan:2011gv,Merkel:2014avp,Merkel:2011ze},
collider~\cite{Aubert:2009cp,Curtin:2013fra,Lees:2014xha,Ablikim:2017aab,Aaij:2017rft,Anastasi:2015qla},
and rare-meson-decay \cite{Bernardi:1985ny,MeijerDrees:1992kd,Archilli:2011zc,Gninenko:2011uv,Babusci:2012cr,Adlarson:2013eza,Agakishiev:2013fwl,Adare:2014mgk,Batley:2015lha,KLOE:2016lwm}
experiments,
and on invisible \aprime decays in Refs.~\cite{Essig:2013vha,Davoudiasl:2014kua,Banerjee:2016tad,Lees:2017lec,Adler:2001xv,Adler:2004hp,Artamonov:2009sz,Fayet:2006sp,Fayet:2007ua,Fox:2011fx}.
Many ideas have been proposed to further explore the \maeps parameter space in the future~\cite{Essig:2010xa,
Freytsis:2009bh,Balewski:2013oza,
Wojtsekhowski:2012zq,
Beranek:2013yqa,
Echenard:2014lma,
Battaglieri:2014hga,
Alekhin:2015byh,
Gardner:2015wea,
Ilten:2015hya,
Curtin:2014cca,
He:2017ord,Kozaczuk:2017per,
Ilten:2016tkc,
Feng:2017uoz}.

Both existing and proposed searches for dark photons provide serendipitous discovery potential for other types of vector particles.
Therefore, interpreting these results within the context of a more generic model is well motivated.
In this article, we develop a framework for recasting searches for massive vector particles from one model to another, which includes a data-driven method for determining hadronic decay rates.
We demonstrate our approach by recasting the existing constraints on dark photons; however, we stress that
our approach can easily be applied to any massive gauge boson with vector couplings to the SM fermions.

A variety of production mechanisms have been used in dark-photon searches, which can be categorized as follows:
\begin{list}{\textbullet}{\leftmargin=1em}
\item bremsstrahlung, $eZ\to eZ\aprime$ and $pZ\to pZ\aprime$, using electron and proton beams incident on fixed nuclear targets of charge $Z$;
\item annihilation, $e^+e^-\to\aprime\gamma$, at $e^+e^-$ colliders;
\item Drell-Yan~(DY), $q\bar{q}\to\aprime$, both at hadron colliders and at proton-beam fixed-target experiments;
\item meson decays, {\em e.g.}\ $\pi^0\to\aprime\gamma$, $\eta\to\aprime\gamma$, $\omega\to\aprime\pi^0$, and $\phi\to\aprime\eta$;
\item and $V\to\aprime$ mixing, where $V=\omega,\rho,\phi$ denotes the QCD vector mesons.
\end{list}
Proposed future searches largely exploit the same production mechanisms, though some plan on using positron beams incident on fixed targets for annihilation~\cite{Wojtsekhowski:2012zq,Raggi:2014zpa,Alexander:2017rfd}
or additional meson decays such as $D^*\to D^0\aprime$\,\cite{Ilten:2015hya}.
Dark photons have been searched for using the following techniques:
\begin{list}{\textbullet}{\leftmargin=1em}
\item by performing bump hunts in invariant mass spectra using the visible decays $\aprime\to\ell^+\ell^-$ and $\aprime\to h^+h^-$, where thus far $\ell=e,\mu$ and $h=\pi$ have been used;
\item by searching for visible displaced \aprime decays, which has been done both at beam dumps and at colliders using secondary vertices;
\item and by performing bump hunts in missing mass spectra, which requires the initial state to be known and any visible component of the final state to be detected, providing sensitivity to invisible \aprime decays.
\end{list}
While the production mechanisms and search strategies employed were chosen to achieve the best possible sensitivity to dark photons, each also provides sensitivity to other types of hypothesized vector particles.

The remainder of this article is organized as follows.
Section~\ref{sec:framework} develops the framework required to recast these searches, which includes a novel and robust method for determining the hadronic decay rates for \gev-scale bosons.
We apply our framework to three models in Sec.~\ref{sec:models}: a vector that couples to the \BL current, a leptophobic $B$ boson that couples directly to baryon number and to leptons via $B$--$\gamma$ kinetic mixing, and on a vector that mediates a protophobic force.
Finally, summary and discussion are provided in Sec.~\ref{sec:sum}.
{\em N.b.}, all information required to recast dark photon searches to any vector model, including software to perform any such recasting, is provided at \git.

\section{Generic Vector Boson Model}
\label{sec:framework}

In this section, we consider a generic model that couples a vector boson $X$ to SM fermions, $f$, and to invisible dark-sector particles, $\chi$, according to
\begin{equation}
  \label{eq:L}
  \cL \subset  g_X \sum_f x_{f} \bar{f} \gamma^{\mu} f X_{\mu} + \sum_{\chi} \cL_{X\chi\bar{\chi}} \, ,
\end{equation}
where $g_X x_f$ is the coupling strength to fermion $f$, and the form of the $X\chi\bar{\chi}$ interaction does not need to be specified.\footnote{This model is flavor-conserving due to its diagonal couplings. Of course, one could also consider flavor-violating $X$ couplings; however, in such cases, the constraints from studies of flavor-changing neutral currents are much stronger than those from \aprime searches. Furthermore, we only consider real $x_f$ for similar reasons, making this a $CP$-conserving model as well.}
For example, in the minimal \aprime scenario, where the \aprime coupling to SM fermions arises due to $\gamma$--\aprime kinetic mixing, $g_X = \varepsilon e$, $x_{\ell} = -1$, $x_{\nu}=0$, and $x_q = 2/3$ or $-1/3$.
The \aprime also has a model-dependent coupling to the weak $Z$ current that scales as $\cO(m^2_{\aprime}/m^2_Z)$, see {\em e.g.}\ Ref.~\cite{Barello:2015bhq}.
For $m_{\aprime}>10\gev$, we adopt the model of Refs.~\cite{Cassel:2009pu,Cline:2014dwa}.
The \aprime decays visibly if $\ma < 2m_{\chi}$ for all $\chi$, and predominantly invisibly otherwise.
The more general model has 14 parameters: the 12 fermion couplings, the $X$ boson mass, $m_X$, and its decay branching fraction into invisible dark-sector final states.

Recasting a dark photon search that used the final state $\cF$ involves solving the following equation for each $m_X = \ma$:
\begin{equation}
  \label{eq:xtoa}
  \sigma_X \cB_{X\to\cF}\, \epsilon(\tau_X)  = \sigma_{\aprime} \cB_{\aprime \to \cF}\, \epsilon(\tau_{\aprime}) \, ,
\end{equation}
where $\sigma_{X,\aprime}$ denotes the production cross section,
$\mathcal{B}_{X,\aprime\to\cF}$ is the decay branching fraction,
and $\epsilon$ is the detector efficiency, whose lifetime dependence is made explicit.
From Eq.~\eqref{eq:xtoa}, one can see that what is needed are the ratios $\sigma_X / \sigma_{\aprime}$, $\cB_{X \to \cF}/\cB_{\aprime \to \cF}$, and $\epsilon(\tau_{X})/ \epsilon(\tau_{\aprime})$.
{\em N.b.}, in models where the $X$ couples to an anomalous SM current, there are additional strong constraints from the $B_{u,d}\to KX$, $Z\to\gamma X$, and $K\to\pi X$ processes, which arise due to the enhanced production rates of the longitudinal $X$ mode~\cite{Dror:2017nsg,Dror:2017ehi,Ismail:2017fgq}.

\subsection{$X$ production}
\label{sec:xprod}

The ratio of production cross sections for both electron-beam bremsstrahlung and $e^+e^-$ annihilation is
\begin{equation}
  \frac{\sigma_{eZ\to eZX}}{\sigma_{eZ\to eZ\aprime}} = \frac{\sigma_{e^+e^-\to X\gamma}}{\sigma_{e^+e^-\to\aprime\gamma}} = \frac{(g_X x_e)^2}{(\varepsilon e)^2}.
\end{equation}
For proton-beam bremsstrahlung the situation is more complicated, but to a good approximation the ratio can be taken to be
\begin{equation}
  \frac{\sigma_{pZ\to pZX}}{\sigma_{pZ\to pZ\aprime}} \approx  \frac{g_X^2(2x_u+x_d)^2}{(\varepsilon e)^2} \, ,
\end{equation}
since only sub-GeV masses have been probed using this production mechanism.
The ratio of DY production cross sections involves a sum over quark flavors, $q_i$, and is given by
\begin{equation}
  	\frac{\sigma_{{\rm DY}\to X}}{\sigma_{{\rm DY}\to\aprime}}
= 	\sum_{q_i} \left[\frac{\sigma_{q_i\bar{q}_i \to \gamma^*}(m)}{\sigma_{{\rm DY} \to \gamma^*}(m)} \right]
	\left[\frac{\sigma_{q_i\bar{q}_i\to X}}{\sigma_{q_i\bar{q}_i\to \aprime}} \right] \, ,
\end{equation}
where the first term in the sum is the mass-dependent fraction of the SM DY production attributed to each flavor, and the second term is the contribution from each subprocess
\begin{equation}
  \label{eq:dy}
  	\frac{\sigma_{q_i\bar{q}_i\to X}}{\sigma_{q_i\bar{q}_i\to\aprime}} = \frac{9(g_X x_{q_i})^2}{(\varepsilon e)^2} \times
	\begin{cases}
    		\frac{1}{4} & \text{for } q_i = u,c, \\
    		1 & \text{for } q_i = d,s,b.
  	\end{cases}
\end{equation}
For $m_X \gtrsim 10\gev$, the model-dependent mixing with the $Z$ must be accounted for in Eq.~\eqref{eq:dy}.
Furthermore, the value of $e$ should be evaluated at the proper mass scale, though this is a small effect below $m_Z$.
Determining the fraction of SM DY production attributed to each flavor requires knowledge of the parton distribution functions of the proton, though the uncertainties that arise due to limitations in this knowledge largely cancel in the ratios.

Following Ref.~\cite{Tulin:2014tya}, we calculate meson-decay ratios using the hidden local symmetries framework of vector meson dominance (VMD)~\cite{Fujiwara:1984mp}, which is successful at predicting low-energy SM observables.\footnote{The VMD approach accurately predicts many observables at the 10--20\% level, {\em e.g.}, the width of the $\omega$ meson~\cite{Fujiwara:1984mp}. Therefore, we expect that the uncertainty of using VMD and $U(3)$ quark symmetry is $\approx 20\%$.}
In this effective theory, external gauge fields---including the SM photon---couple to quarks via mixing with the QCD vector mesons.
The ratio of the widths for producing the $X$ and \aprime in decays of the form $V\to XP$, where $V$ and $P$ denote vector and pseudoscalar mesons, respectively, is given by
\begin{equation}
  \label{eq:vmd_main}
	\frac{\Gamma_{V\to X P}}{\Gamma_{V \to \aprime P}}
  = 	\frac{g_X^2}{(\varepsilon e)^2} \frac{\left|\sum_{V^{\prime}} {\rm Tr}[T_V T_P T_{V^{\prime}}] {\rm Tr}[T_{V^{\prime}} Q_X] {\rm BW}_{V^{\prime}}(m_X) \right|^2}
  {\left| \sum_{V^{\prime}} {\rm Tr}[T_V T_P T_{V^{\prime}}] {\rm Tr}[T_{V^{\prime}} Q] {\rm BW}_{V^{\prime}}(m_X) \right|^2 } \, ,
\end{equation}
where the sum runs over all possible $VPV'$ vertices.
The quark $U(3)$-charge matrices are
\begin{align}
&	Q = \frac{1}{3}{\rm diag}\{2,-1,-1\} \, , \nonumber \\
&	Q_X = {\rm diag}\{x_u, x_d, x_s \} \, ,
\end{align}
and the relevant meson generators, $T_{V,P}$,
and the VMD Breit-Wigner form factors, ${\rm BW}_V(m)$, are detailed in Appendix~\ref{app:vmd}.
When considering $V$ and $P$ from the lowest-lying nonets, where VMD is valid, this reduces to
\begin{equation}
  \label{eq:VVP}
  \frac{\Gamma_{V\to XP}}{\Gamma_{V\to\aprime P}} = \left(\frac{g_X}{\varepsilon e}\right)^2 \frac{\{{\rm Tr}[T_{V^{\prime}}Q_X] \}^2}{ \{{\rm Tr}[T_{V^{\prime}}Q] \}^2},
\end{equation}
where $V^{\prime}$ is chosen such that the process $V \to V^{\prime}P$ is $SU(3)$ allowed, {\em e.g.}\ $\omega \to \omega\eta$ and $\omega\to\rho\pi^0$ are allowed, whereas  $\omega \to \rho\eta$ and $\omega\to\omega\pi^0$ are not.
The ratio of widths for $P \to X\gamma$ and $P \to \aprime \gamma$ decays satisfies a similar expression:
\begin{equation}
  \label{eq:PVG}
\frac{\Gamma_{P\to X\gamma}}{\Gamma_{P\to\aprime \gamma}} = \left(\frac{g_X}{\varepsilon e}\right)^2
\frac{\left| \sum_V {\rm Tr}[T_P Q T_V] {\rm Tr}[T_V Q_X] {\rm BW}_V(m) \right|^2}{\left|  \sum_V {\rm Tr}[T_P Q T_V] {\rm Tr}[T_V Q] {\rm BW}_V(m)  \right|^2},
\end{equation}
which cannot be reduced into as simple a form due to the fact that multiple terms in the sum over $V$ contribute.
Finally, the ratio of production cross sections due to the $X$ mixing with the QCD vector mesons is
\begin{equation}
  	\frac{\sigma_{V\to X}}{\sigma_{V\to\aprime}}
= 	\frac{g_X^2}{(\varepsilon e)^2} \times
  	\begin{cases}
    		(x_u-x_d)^2 & \text{for } V = \rho, \\
   		 9(x_u+x_d)^2 & \text{for } V = \omega, \\
    		9x_s^2 & \text{for } V = \phi,
  	\end{cases}
\end{equation}
which is also calculated using VMD.
This approach ignores potential interference between the $\rho$, $\omega$, and $\phi$ production amplitudes.\footnote{Including such interference is trivial if the relative phases of the amplitudes are known; however, this production mechanism is only important in hadronic environments, where these phases are generally not known and where interference effects are expected to be negligible.}

The sensitivity in many dark photon searches is predominantly due to a single production mechanism at each mass.
In such cases, the ratio $\sigma_X / \sigma_{\aprime}$ is obtained directly from one of the ratios provided in this subsection.
When more than one production mechanism is relevant, the cross-section ratio is
\begin{equation}
  	\frac{\sigma_X}{\sigma_{\aprime}}
= 	\sum_{i} \left[\frac{\sigma_{\aprime}^i}{\sigma_{\aprime}} \right] \left[\frac{\sigma_X^i}{\sigma_{\aprime}^i} \right],
\end{equation}
where a Monte Carlo event generator can be used to estimate the relative importance of each production mechanism.

\subsection{$X$ decays}
\label{sec:xdec}

The $X$ boson is assumed to decay predominantly into invisible dark-sector final states if kinematically allowed,
and into SM final states otherwise.
The partial width of the decay $X\to f\bar{f}$ is given by
\begin{equation}
  	\label{eq:G}
	\Gamma_{\!X\to f\bar{f}} = \frac{\mathcal{C}_f(g_X x_f)^2}{12\pi}m_X\left(1+2\frac{m_f^2}{m_X^2}\right) \sqrt{1-4\frac{m_f^2}{m_X^2}} \, ,
\end{equation}
where $\mathcal{C}_f = 1$ for $\ell^+\ell^-$, 3 for $q\bar{q}$, and $1/2$ for $\nu\bar{\nu}$; however, for masses $\lesssim 2$\,GeV, we do not expect
to obtain a reliable prediction for $\Gamma_{X\to{\rm hadrons}}$ by summing the $q\bar{q}$ contributions from Eq.~\eqref{eq:G}.
Because the \aprime couples to the EM current, its decay rate into hadrons is simply
\begin{equation}
  \Gamma_{\aprime\to{\rm hadrons}} = \Gamma_{\aprime\to\mu^+\mu^-}\mathcal{R}_{\mu}(\ma),
\end{equation}
where $\mathcal{R}_{\mu} \equiv \sigma(e^+e^-\!\to\!{\rm hadrons})/\sigma(e^+e^-\!\to\!\mu^+\mu^-)$ is known experimentally~\cite{Patrignani:2016xqp}.
{\em N.b.}, this expression already accounts for $\aprime \to V$ mixing.

The VMD approach can be used to estimate $\Gamma_{X\to\mathcal{F}}$ for specific hadronic final states when $m_X \lesssim  m_{\phi}$, but not in the region from 1 to 2\,GeV.
To obtain reliable predictions for all masses, we have instead developed a data-driven approach based on measured $e^+e^-\to\mathcal{F}$ cross sections.
First, we normalize each of the most important $e^+e^-\to\cF$ hadronic cross sections at low mass to that of $e^+e^- \to \mu^+\mu^-$
\begin{equation}
  	\label{eq:Rfit}
  	\cR_{\mu}^{\cF}(m)
	\equiv
	\frac{\sigma_{e^+e^- \to \cF}}{\sigma_{e^+e^- \to \mu^+\mu^-}}
= 	\frac{9}{\alpha^2_{\rm EM}} |\cA_{\cF}(m)|^2 \, ,
\end{equation}
where $m$ is the $e^+e^-$ invariant mass.
Each $\cF$-dependent amplitude is taken to be the sum of a real function $f_{\mathcal{F}}(m)$, which accounts for $V^*$ components, and contributions from $V=\rho,\omega,\phi$ as
\begin{equation}
    	\label{eq:Rfit2}
  	\cA_{\cF}(m)
= 	f_{\cF}(m) \pm \sum_{V} \mathcal{A}_{\cF}^V(m) \, ,
\end{equation}
where the minus sign applies only to $V=\phi$, arising from ${\rm Tr}[T_{\phi}Q] < 0$, and the $\mathcal{A}_{\cF}^V(m)$ amplitudes have Breit-Wigner forms which are provided in Appendix~\ref{app:xtohad}.
Taking $f_{\cF}(m)$ to be real corresponds to the assumption that the only relevant interference effects between $V^*$ and $V$ occur far from the $V^*$ poles, which is demonstrated to be a good approximation in Appendix~\ref{app:xtohad}.

\begin{figure*}[p!]
  \centering
  \includegraphics[width=0.99\textwidth]{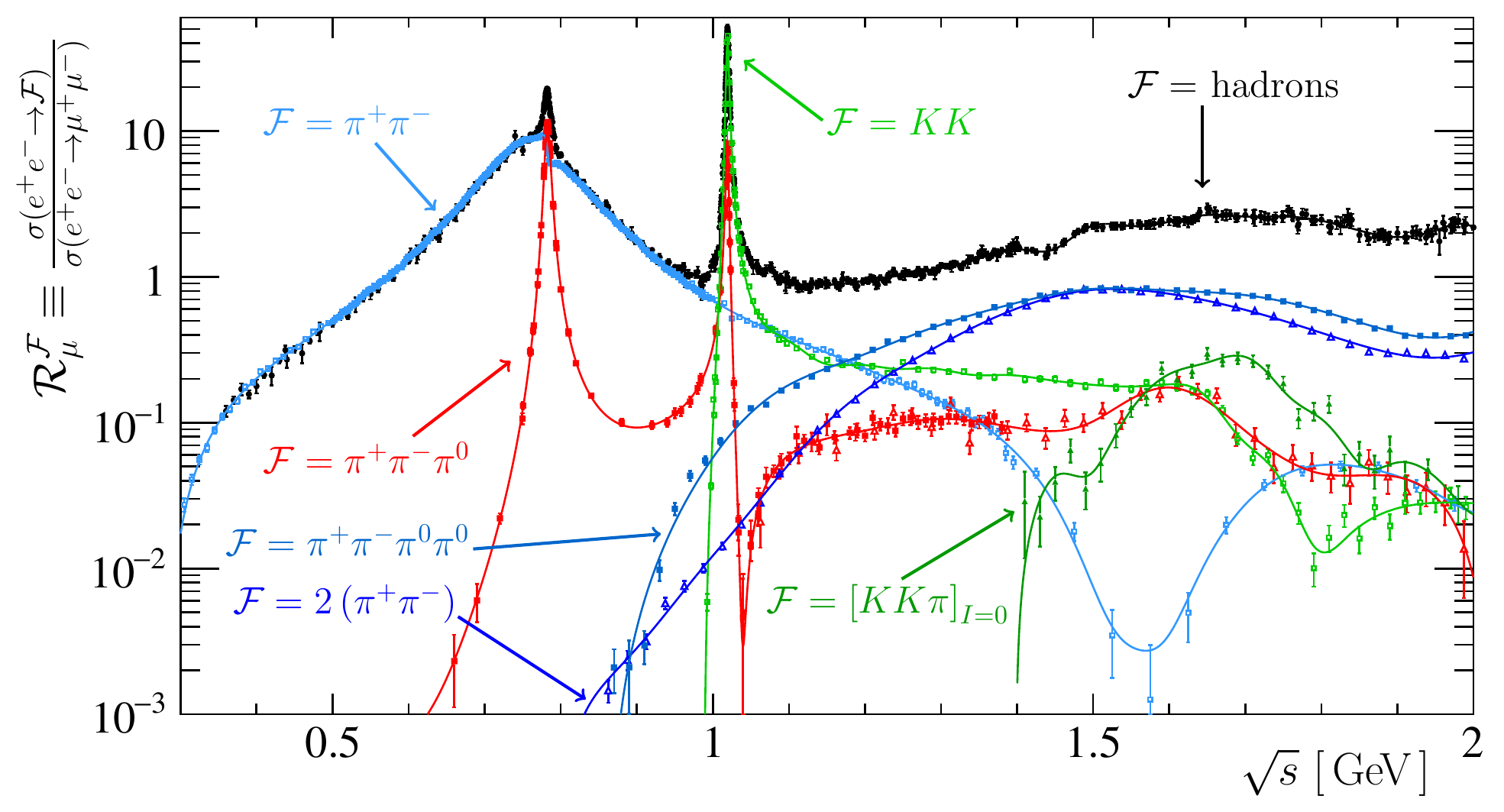}
  \caption{
  Data used to determine the hadronic decay rates from:
  the PDG, for the total rate to hadrons\,\cite{Patrignani:2016xqp};
  BaBar, for $\pi^+\pi^-$\,\cite{Lees:2012cj},
  high-mass $\pi^+\pi^-\pi^0$\,\cite{Aubert:2004kj} (displayed as open triangles),
  $KK\equiv K^+K^- + K_SK_L$\,\cite{Lees:2013gzt},
  $\left[KK\pi\right]_{I=0}$\,\cite{Aubert:2007ym} ({\em i.e.}\ the isoscalar component of the $KK\pi$ final state),
  $2(\pi^+\pi^-)$\,\cite{Lees:2012cr}, and
  $\pi^+\pi^-\pi^0\pi^0$\,\cite{TheBaBar:2017vzo};
  and from SND, the low-mass $\pi^+\pi^-\pi^0$\,\cite{Achasov:2003ir,Achasov:2002ud} (displayed as filled squares).
  See text for discussion on the solid lines.
  }
  \label{fig:rmu_data}
\end{figure*}

The six most important hadronic contributions to $\cR_{\mu}$ at low mass are fitted using Eq.~\eqref{eq:Rfit}, where each $f_{\mathcal{F}}(m)$ is taken to be a bicubic spline with knots every 50\mev whose values are varied to achieve the best description of the data.
Figure~\ref{fig:rmu_data} shows that these fits describe all data samples well.
For $m \lesssim 0.5\gev$, the $\pi^0\gamma$ final state is the dominant $\omega$ decay mode.
Due to a lack of $e^+e^-\to\pi^0\gamma$ data where this decay is important, we instead calculate this contribution assuming it comes entirely from the $\omega$; {\em i.e.}\ this contribution is estimated as above, but with $f_{\pi^0\gamma}(m)=0$ and only using $V = \omega$.

Based on these fits, we are able to decompose $e^+e^-\!\!\to$\,hadrons into $\rho$-like, $\omega$-like, and $\phi$-like contributions, which are discussed in detail in Appendix~\ref{app:xtohad} and shown in Fig.~\ref{fig:rmu_data2}.
Each of these contributions is within 20\% of its leading order~(LO) perturbative value for $m \gtrsim 1.5\gev$, as is $\mathcal{R}_{\mu}$ itself, justifying the use of LO perturbative $\Gamma_{X \to {\rm hadrons}}$ values above 2\gev.
Using these $\rho$-like, $\omega$-like, and $\phi$-like models, we can estimate $\Gamma_{X \to {\rm hadrons}}$ for any low-mass $X$ model  from
\begin{equation}
	\label{eq:GammaXhad}
  	\Gamma_{X \to {\rm hadrons}}
\!=\! 	\frac{g_X^2 m_X}{12\pi} \left[ \sum_V \mathcal{R}_{X}^{V}(m_X) \! + \! \mathcal{R}_{X}^{\omega\textrm{-}\phi}(m_X)\right],
\end{equation}
where
\begin{eqnarray}
	\label{eq:RX}
	\mathcal{R}_{X}^{\rho}(m) &=& \{2{\rm Tr}[T_{\rho}Q_X]\}^2 \mathcal{R}_{\mu}^{\rho}(m)  \, , \nonumber \\
	\mathcal{R}_{X}^{\omega}(m) &=&  \{6{\rm Tr}[T_{\omega}Q_X]\}^2 \mathcal{R}_{\mu}^{\omega}(m) \,  , \\
	\mathcal{R}_{X}^{\phi}(m) &=& \{3\sqrt{2}{\rm Tr}[T_{\phi}Q_X]\}^2 \mathcal{R}_{\mu}^{\phi}(m) \, . \nonumber
\end{eqnarray}
The final term in Eq.~\eqref{eq:GammaXhad} accounts for interference between the $\omega$-like and $\phi$-like contributions to the $\pi^+\pi^-\pi^0$ final state and is given by
\begin{equation}
	\label{eq:RXint}
	\mathcal{R}_{X}^{\omega\textrm{-}\phi}(m) = 	36\sqrt{2} {\rm Tr}[T_{\omega}Q_X] {\rm Tr}[T_{\phi}Q_X] \, \Re\big\{ \mathcal{A}^{\phi}_{3\pi}(m) \left[f_{3\pi}(m) + \mathcal{A}^{\omega}_{3\pi}(m) \right]^*\big\}.
\end{equation}
All other interference effects between the $\rho$-like, $\omega$-like, and $\phi$-like contributions are assumed to be negligible.\footnote{The numerical values of the $\cR^\cF_X$ functions defined in Eqs.~\eqref{eq:RX}--\eqref{eq:RXint} are provided at \git.}

\begin{figure*}[p!]
  \centering
  \includegraphics[width=0.99\textwidth]{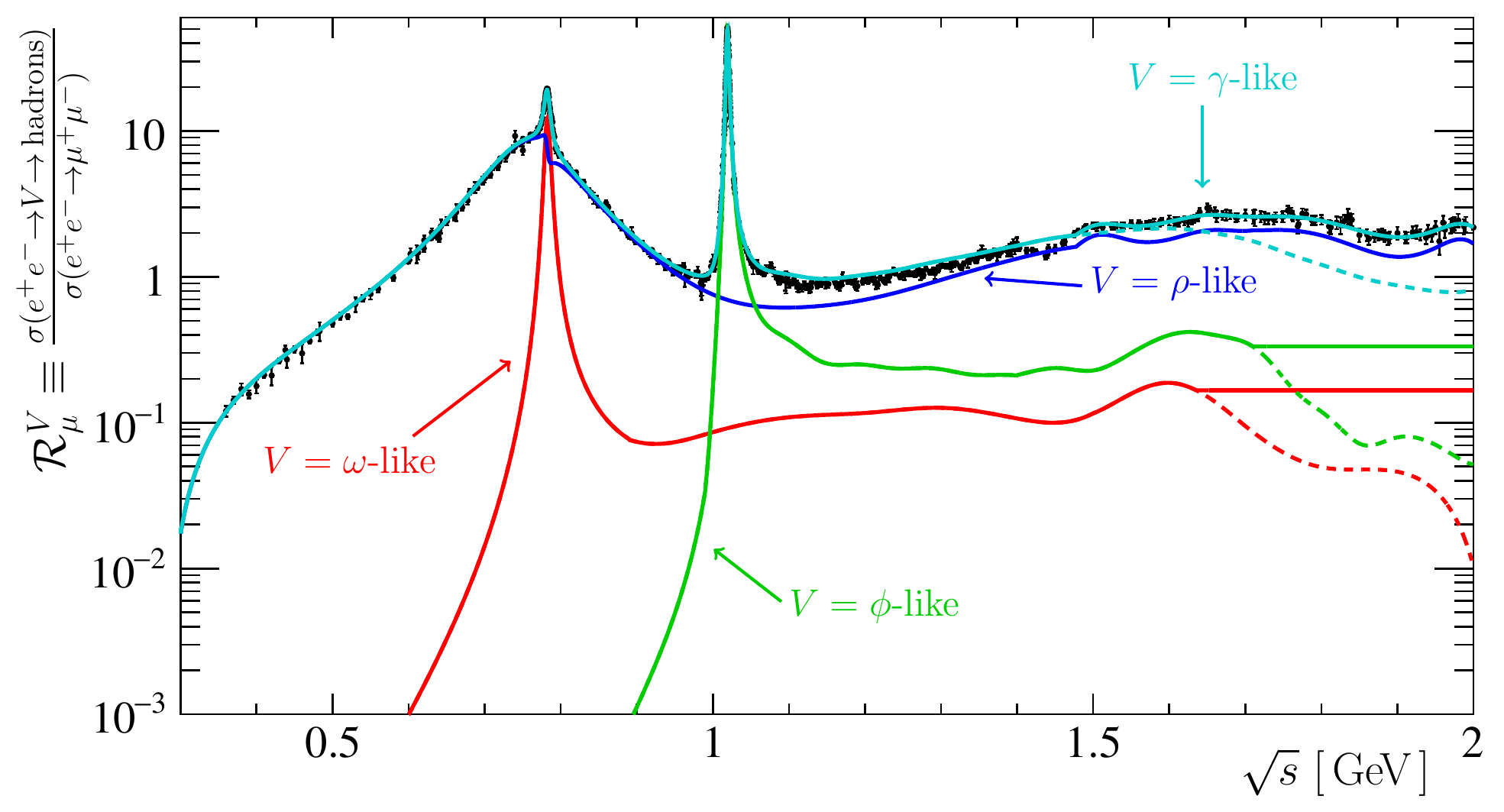}
  \caption{
  Decomposition of $e^+e^-\!\!\to$\,hadrons, which is of course $\gamma$-like, into
  ($\rho$-like) $(u\bar{u}-d\bar{d})/\sqrt{2}$,
  ($\omega$-like) $(u\bar{u}+d\bar{d})/\sqrt{2}$,
  and
  ($\phi$-like) $s\bar{s}$ contributions.
  See Appendix~\ref{app:xtohad} for detailed discussion on the derivation of these curves and on the meaning of the dashed lines.
  }
  \label{fig:rmu_data2}
\end{figure*}

We reiterate that the approach developed here, specifically Eq.~\eqref{eq:GammaXhad}, can be used to obtain $\Gamma_{X \to {\rm hadrons}}$ for any vector model at low mass, where all that is needed as input are the couplings of the $X$ to the $u$, $d$, and $s$ quarks.
Our approach reproduces $\Gamma_{\aprime\to{\rm hadrons}}$ by construction when the model parameters are chosen to be those of the dark photon.
While our method invokes a few mild assumptions, this is unavoidable and we believe that the approach developed here is the most robust method for determining the hadronic decay rate of a low-mass vector boson.

\subsection{Efficiency ratios}
\label{sec:effratios}

The ratio of detector efficiencies for the $X$ relative to the \aprime is taken to be unity for invisible searches.
Searches for visible prompt \aprime decays also have the same efficiency for the $X$, provided that $\tau_X$ is smaller than the detector decay-time resolution.
This is not the case for all models; therefore, lifetime-dependent efficiency effects must be considered even in prompt searches.
All existing prompt \aprime searches had $\epsilon(\tau_{\aprime}) \approx 1$ which gives
\begin{equation}
  \label{eq:prompt_eff}
	\frac{\epsilon(\tau_X)}{\epsilon(\tau_{\aprime})} \approx 1 - e^{-\tilde{t}/\tau_X} \, ,
\end{equation}
where $\tilde{t}$ denotes the largest proper decay time that an $X$ boson could have and still satisfy the prompt \aprime search selection criteria.
The experiment-dependent $\tilde{t}$ values are provided in Appendix~\ref{app:exp}.

The efficiency ratios are more complicated in searches for long-lived bosons.
The recent LHCb search~\cite{Aaij:2017rft} for $\aprime \to \mu^+\mu^-$ published not only the \aprime exclusion regions, but also the ratio, $r^{\rm ul}_{\rm ex}$, of the upper limit on the observed \aprime yield relative to the expected number of observed \aprime decays at each $[\ma, \varepsilon^2]$.
For the \aprime, regions with $r^{\rm ul}_{\rm ex} < 1$ are excluded.
This facilitates recasting the results for each $\tau_X = \tau_{\aprime}$, where the ratio of efficiencies is again unity.
Regions with
\begin{equation}
  \label{eq:lhcb_displ_recast}
	\left[  r^{\rm ul}_{\rm ex}(\ma,\varepsilon^2) \frac{\sigma_{\aprime} \mathcal{B}_{\aprime \to \mathcal{F}}}{\sigma_{X} \mathcal{B}_{X \to \mathcal{F}}} \right]_{\tau_X = \tau_{\aprime}} \!\! < \,\, 1 \, ,
\end{equation}
are excluded for the $X$.
We encourage future beam dump and displaced-vertex searches to also publish results in this way (or similarly, $r^{\rm ul}_{\rm ex}$ at each $[\ma, \tau_{\aprime}]$), as it makes recasting the results trivial.
{\em N.b.}, the LHCb sensitivity for some models extends to $\tau_X$ values for which LHCb does not report results, though these regions are easily handled as discussed in Appendix~\ref{app:exp.lhcb}.

\clearpage

The published information for constraints placed on dark photons from beam-dump experiments is not sufficient to rigorously recast the results for other models.
In principle, the Monte Carlo studies need to be redone, and the $r^{\rm ul}_{\rm ex}$ values extracted for each $[\ma,\varepsilon^2]$ as was done at LHCb~\cite{Aaij:2017rft}.
That is beyond the scope of this project.
Instead, we set approximate limits by defining an effective proper-time fiducial decay region of $[\tilde{t}_0,\tilde{t}_1]$ for each experiment, where
$\tilde{t}_1$ can be written in terms of the lengths of the decay volume, $L_{\rm dec}$, and shielding, $L_{\rm sh}$, as
\begin{equation}
  	\tilde{t}_1 = \tilde{t}_0 (1 + L_{\rm dec}/L_{\rm sh}) \, .
\end{equation}
This approach ignores the kinematical spread of the production momentum spectra and the dependence of the efficiency on the location of the decay within the decay volume, though a proper treatment amounts to an $\mathcal{O}(1)$ correction to limits that cover several orders of magnitude for the existing beam-dump results.
The probability that a particle with lifetime $\tau$ decays within this fiducial region is given by
\begin{equation}
    \label{eq:eff_dump}
  \epsilon(\tau) = e^{-\tilde{t}_0 / \tau } - e^{-\tilde{t}_1 / \tau } \, .
\end{equation}
The values for $\tilde{t}_0$ and $\tilde{t}_1$ are obtained at each mass from the \aprime limits $[\varepsilon_{\rm min},\varepsilon_{\rm max}]$ by solving
\begin{equation}
  	\label{eq:t0}
  	\varepsilon^2_{\rm max}   \epsilon[\tau_{\aprime}(\varepsilon^2_{\rm max})]
= 	\varepsilon^2_{\rm min}   \epsilon[\tau_{\aprime}(\varepsilon^2_{\rm min})] \, ,
\end{equation}
which arises from the fact that the upper limit on observed signal decays is independent of decay time, {\em i.e.}\ the experimental upper limits placed on observed signal decays do not depend on the decay time.

We provide here some simple heuristics that give nearly identical results to the more involved approach described above, provided that the beam-dump experiment is sensitive to the $X$ model being considered at a given mass.
For the upper edge of a long-lived \aprime exclusion region, the \aprime lifetime is much smaller than the minimum proper decay time required to enter the beam-dump fiducial region.
This means that the efficiency is exponentially suppressed\,(enhanced) for $\tau_X < \tau_{\aprime}$\,$(\tau_X > \tau_{\aprime})$, resulting in the upper edge of the exclusion region for the $X$ occurring at the $g_X$ value where
\begin{equation}
	\tau_X(g_X^{\rm max}) \approx \tau_{\aprime}(\varepsilon^{\rm max}) \, .
\end{equation}
The lower eddge of the \aprime exclusion region is typically where the \aprime lifetime is much larger than the maximum proper decay time required to decay before exiting the fiducial region.
In this regime, the ratio of efficiencies is  just the ratio of the lifetimes, and the lower edge of the $X$ exclusion region occurs where
\begin{equation}
	\left[\frac{\sigma_X \mathcal{B}_{X\to\mathcal{F}}}{\tau_X}\right]_{g_X^{\rm min}}
	\approx
	\left[\frac{\sigma_{\aprime} \mathcal{B}_{\aprime\to\mathcal{F}}}{\tau_{\aprime}}\right]_{\varepsilon^{\rm min}}
\end{equation}
is satisfied.
We do not use these heuristics to obtain the results presented in Sec.~\ref{sec:models}, though they do give nearly identical results except near the high-mass edges of the beam-dump exclusion regions, where the large-lifetime approximation is no longer valid at the lower edges of the \aprime exclusion regions.

\section{Example Models}
\label{sec:models}

We now use the framework developed in the previous section to recast existing dark photon searches to obtain constraints on the following models: a vector that couples to the \BL current, a leptophobic $B$ boson that couples directly to baryon number and to leptons via $B$--$\gamma$ kinetic mixing, and on a vector that mediates a protophobic force~\cite{Feng:2016ysn}.
The fermionic couplings of each of these models are provided in Table~\ref{tab:models}.
Using these couplings and the results of Sec.~\ref{sec:xprod}---including the work in Appendix~\ref{app:vmd}---it is straightforward to obtain all of the necessary $\sigma_X / \sigma_{\aprime}$ ratios, which are summarized in Tables~\ref{tab:prod} and \ref{tab:meson_decays}.
First, we will recast the \aprime searches assuming $\mathcal{B}(X\to\chi\bar{\chi})=0$ for each of these three models, followed by recasting each of them under the assumption $\mathcal{B}(X\to\chi\bar{\chi})\approx 1$.
{\em N.b.}, we do not consider astrophysical constraints in either case (see, {\em e.g.}, Ref.~\cite{Heeck:2014zfa}).

\begin{table*}[t]
  \begin{center}
    \caption{\label{tab:models} Couplings to SM fermions for the models studied in Sec.~\ref{sec:models}.}
    {\renewcommand{\arraystretch}{2}
    \begin{tabular}{c|cccc}
      Coupling & \phantom{xxxxx}\aprime\phantom{xxx} & \phantom{xxxx}\BL\phantom{xx} & \phantom{xxxxx}$B$\phantom{xxxxx}  & Protophobic \\
      \hline
      $g_X$ & $\phantom{-}\varepsilon e$ & $\phantom{-}g_{\BL}$ & $g_B$ & $\phantom{-}g_{\pphob}$ \\
      $\displaystyle x_{u,c,t}$ & $\phantom{-}\displaystyle\frac{2}{3}$ & $\phantom{-}\displaystyle\frac{1}{3}$ & $\displaystyle\frac{1}{3}$ &  $\displaystyle-\frac{1}{3}$  \\
      $\displaystyle x_{d,s,b}$ & $\displaystyle-\frac{1}{3}$ & $\phantom{-}\displaystyle\frac{1}{3}$ & $\displaystyle\frac{1}{3}$ &  $\displaystyle\phantom{-}\frac{2}{3}$  \\
      $\displaystyle x_{e,\mu,\tau}$ & $-1$ & $-1$ & $\displaystyle-\frac{e^2}{(4\pi)^2}$ &  $-1$  \\
      $\displaystyle x_{\nu_e,\nu_{\mu},\nu_{\tau}}$ & $\phantom{-}0$ & $-1$ & 0 &  $\phantom{-}0$  \\
    \end{tabular}\vspace{0.3in}}
  \end{center}
  \end{table*}

\begin{table*}[p!]
  \begin{center}
    \caption{\label{tab:prod} Production rates for the models in Table~\ref{tab:models} relative to those of the dark photon, except for meson-decay rates which are provided in Table~\ref{tab:meson_decays}.}
    {\renewcommand{\arraystretch}{2.7}
    \begin{tabular}{c|ccc}
      Production Mechanism & \phantom{xxx}\BL\phantom{xxx} & \phantom{xxxxx}$B$\phantom{xxxxx}  & Protophobic \\
      \hline
      $\displaystyle\frac{\sigma_{eZ\to eZX}}{\sigma_{eZ\to eZ\aprime}}$  & $\displaystyle\frac{g_{\BL}^2}{(\varepsilon e)^2}$ & $\displaystyle\frac{ e^4 g_{B}^2}{(4\pi)^4 (\varepsilon e)^2}$ & $\displaystyle\frac{g_{\pphob}^2}{(\varepsilon e)^2}$ \\
      $\displaystyle\frac{\sigma_{e^+e^-\to X\gamma}}{\sigma_{e^+e^-\to\aprime\gamma}}$  & $\displaystyle\frac{g_{\BL}^2}{(\varepsilon e)^2}$ & $\displaystyle\frac{e^4 g_{B}^2}{(4\pi)^4 (\varepsilon e)^2}$ & $\displaystyle\frac{g_{\pphob}^2}{(\varepsilon e)^2}$ \\
      $\displaystyle\frac{\sigma_{pZ\to pZX}}{\sigma_{pZ\to pZ\aprime}}$ & $\displaystyle\frac{g_{\BL}^2}{(\varepsilon e)^2}$ & $\displaystyle\frac{g_{B}^2}{(\varepsilon e)^2}$ & 0 \\
      $\displaystyle\frac{\sigma_{\{u\bar{u},c\bar{c}\}\to X}}{\sigma_{\{u\bar{u},c\bar{c}\}\to\aprime}}$ & $\displaystyle\frac{g_{\BL}^2}{4(\varepsilon e)^2}$ & $\displaystyle\frac{g_{B}^2}{4(\varepsilon e)^2}$ & $\displaystyle\frac{g_{\pphob}^2}{4(\varepsilon e)^2}$ \\
      $\displaystyle\frac{\sigma_{\{d\bar{d},s\bar{s},b\bar{b}\}\to X}}{\sigma_{\{d\bar{d},s\bar{s},b\bar{b}\}\to\aprime}}$ & $\displaystyle\frac{g_{\BL}^2}{(\varepsilon e)^2}$ & $\displaystyle\frac{g_{B}^2}{(\varepsilon e)^2}$ & $\displaystyle\frac{4 g_{\pphob}^2}{(\varepsilon e)^2}$ \\
      $\displaystyle\frac{\sigma_{\rho\to X}}{\sigma_{\rho\to\aprime}}$ & 0 & 0 & $\displaystyle\frac{g_{\pphob}^2}{(\varepsilon e)^2}$ \\
      $\displaystyle\frac{\sigma_{\omega\to X}}{\sigma_{\omega\to\aprime}}$ & $\displaystyle\frac{4g_{\BL}^2}{(\varepsilon e)^2}$ & $\displaystyle\frac{4g_{B}^2}{(\varepsilon e)^2}$ & $\displaystyle\frac{g_{\pphob}^2}{(\varepsilon e)^2}$ \\
      $\displaystyle\frac{\sigma_{\phi\to X}}{\sigma_{\phi\to\aprime}}$ & $\displaystyle\frac{g_{\BL}^2}{(\varepsilon e)^2}$ & $\displaystyle\frac{g_{B}^2}{(\varepsilon e)^2}$ & $\displaystyle\frac{4g_{\pphob}^2}{(\varepsilon e)^2}$ \\
    \end{tabular}}
  \end{center}
\end{table*}

\begin{table*}[p!]
  \begin{center}
    \caption{\label{tab:meson_decays} Meson-decay rates for the models in Table~\ref{tab:models} relative to those of the dark photon.}
  \end{center} \vspace{-0.3in}
    \hspace{-0.5in}
    {\renewcommand{\arraystretch}{3}
      \begin{tabular}{c|c|c}
        Decay & $B$ ($g_B \to g_{\BL}$ for \BL) & Protophobic \\
        \hline
        $\displaystyle\frac{\Gamma_{\rho^{\pm,0}\to X\pi^{\pm,0}}}{\Gamma_{\rho^{\pm,0}\to\aprime\pi^{\pm,0}}}$ & $\displaystyle\frac{4g_B^2}{(\varepsilon e)^2}$ & $\displaystyle\frac{g_{\pphob}^2}{(\varepsilon e)^2}$ \\
        $\displaystyle\frac{\Gamma_{\rho^0\to X\eta}}{\Gamma_{\rho^0\to\aprime\eta}}$ & 0 & $\displaystyle\frac{g_{\pphob}^2}{(\varepsilon e)^2}$ \\
        $\displaystyle\frac{\Gamma_{\omega\to X\pi^0}}{\Gamma_{\omega\to\aprime\pi^0}}$ & 0 & $\displaystyle\frac{g_{\pphob}^2}{(\varepsilon e)^2}$ \\
        $\displaystyle\frac{\Gamma_{\omega\to X\eta}}{\Gamma_{\omega\to\aprime\eta}}$ & $\displaystyle\frac{4g_B^2}{(\varepsilon e)^2}$ & $\displaystyle\frac{g_{\pphob}^2}{(\varepsilon e)^2}$ \\
        $\displaystyle\frac{\Gamma_{\phi\to X\eta}}{\Gamma_{\phi\to\aprime\eta}}$ & $\displaystyle\frac{g_B^2}{(\varepsilon e)^2}$ & $\displaystyle\frac{4g_{\pphob}^2}{(\varepsilon e)^2}$ \\
        $\displaystyle\frac{\Gamma_{\pi^0\to X\gamma}}{\Gamma_{\pi^0\to\aprime\gamma}}$ & \footnotesize{$\displaystyle\frac{4g_B^2}{(\varepsilon e)^2}\frac{\left|{\rm BW}_{\omega}(m)\right|^2}{\left|{\rm BW}_{\omega}(m) + {\rm BW}_{\rho}(m)\right|^2} \approx \frac{g_B^2}{(\varepsilon e)^2}$} & \footnotesize{$\displaystyle\frac{g_{\pphob}^2}{(\varepsilon e)^2}\frac{\left|{\rm BW}_{\omega}(m) - {\rm BW}_{\rho}(m)\right|^2}{\left|{\rm BW}_{\omega}(m) + {\rm BW}_{\rho}(m)\right|^2} \approx \frac{g_{\pphob}^2 m^4(m_{\omega}^2-m_{\rho}^2)^2}{4(\varepsilon e)^2(m_{\omega}m_{\rho})^4} \approx 0$} \\
        $\displaystyle\frac{\Gamma_{\eta\to X\gamma}}{\Gamma_{\eta\to\aprime\gamma}}$ & \footnotesize{$\displaystyle\frac{4g_B^2}{(\varepsilon e)^2}\frac{\left|{\rm BW}_{\omega}(m) + {\rm BW}_{\phi}(m)\right|^2}{\left|{\rm BW}_{\omega}(m) + 9{\rm BW}_{\rho}(m) - 2{\rm BW}_{\phi}(m)\right|^2}$} & \footnotesize{$\displaystyle\frac{g_{\pphob}^2}{(\varepsilon e)^2}\frac{\left|{\rm BW}_{\omega}(m) - 9{\rm BW}_{\rho}(m) +4{\rm BW}_{\phi}(m)\right|^2}{\left|{\rm BW}_{\omega}(m) + 9{\rm BW}_{\rho}(m) - 2{\rm BW}_{\phi}(m)\right|^2}$} \\
        $\displaystyle\frac{\Gamma_{\eta^{\prime}\to X\gamma}}{\Gamma_{\eta^{\prime}\to\aprime\gamma}}$  & \footnotesize{$\displaystyle\frac{4g_B^2}{(\varepsilon e)^2}\frac{\left|{\rm BW}_{\omega}(m) -2 {\rm BW}_{\phi}(m)\right|^2}{\left|{\rm BW}_{\omega}(m) + 9{\rm BW}_{\rho}(m) +4{\rm BW}_{\phi}(m)\right|^2}$} & \footnotesize{$\displaystyle\frac{g_{\pphob}^2}{(\varepsilon e)^2}\frac{\left|{\rm BW}_{\omega}(m) - 9{\rm BW}_{\rho}(m) -8{\rm BW}_{\phi}(m)\right|^2}{\left|{\rm BW}_{\omega}(m) + 9{\rm BW}_{\rho}(m) +4{\rm BW}_{\phi}(m)\right|^2}$}
      \end{tabular}}
\end{table*}

\subsection{Decays to SM final states}
\label{sec:vissearches}

\begin{figure*}[p!]
  \centering
  \includegraphics[width=0.49\textwidth]{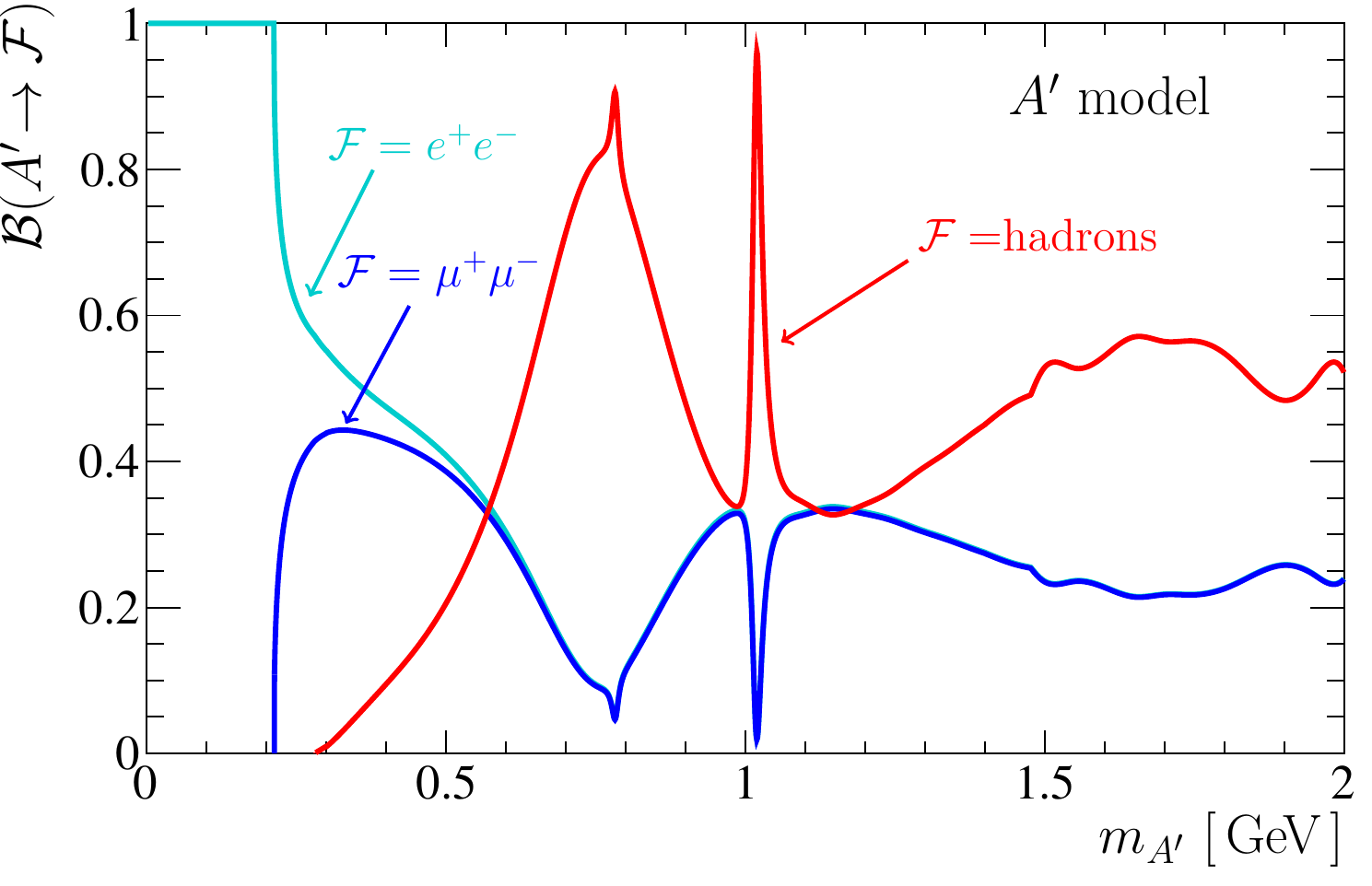}
  \includegraphics[width=0.49\textwidth]{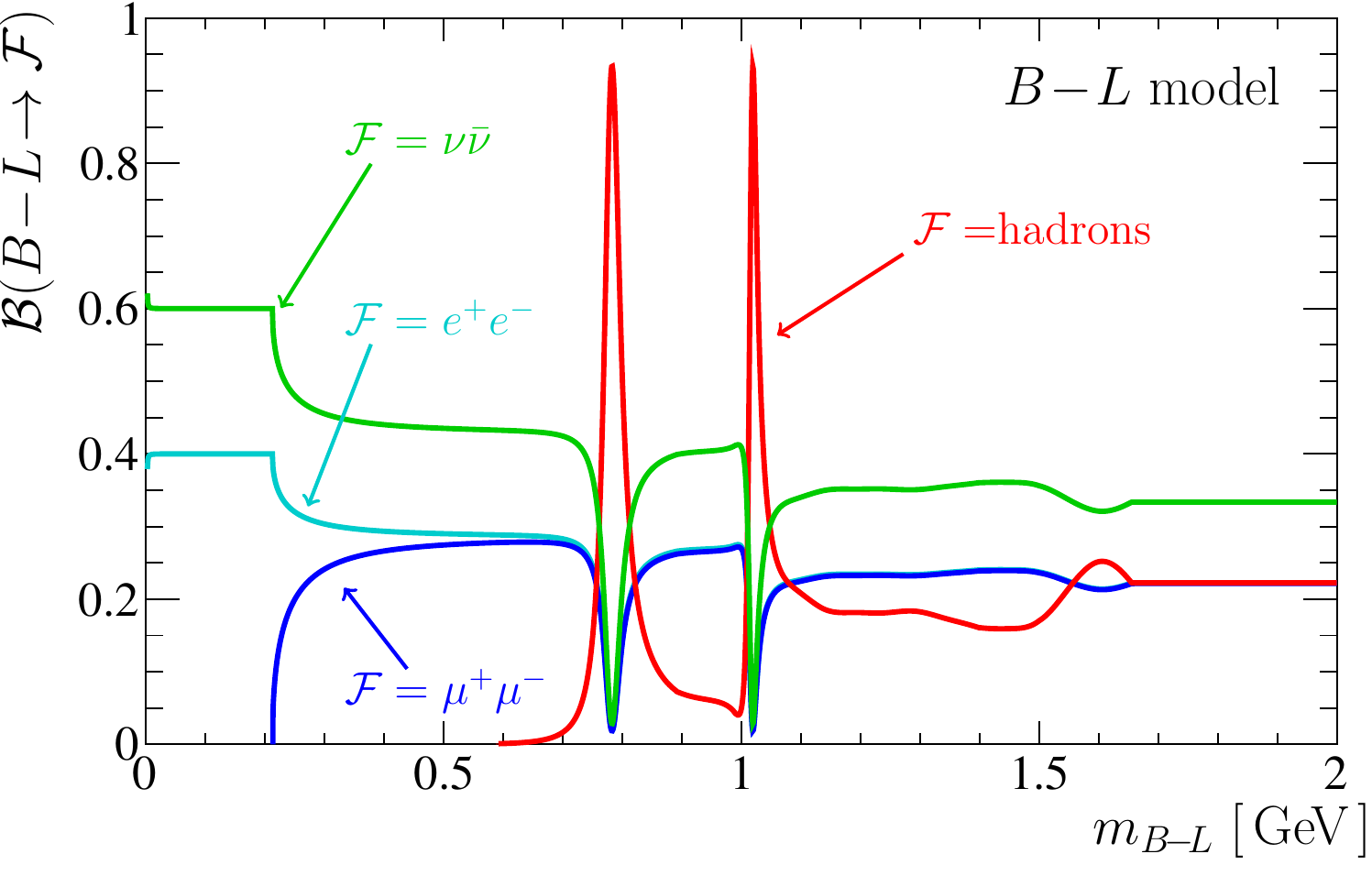}
  \includegraphics[width=0.49\textwidth]{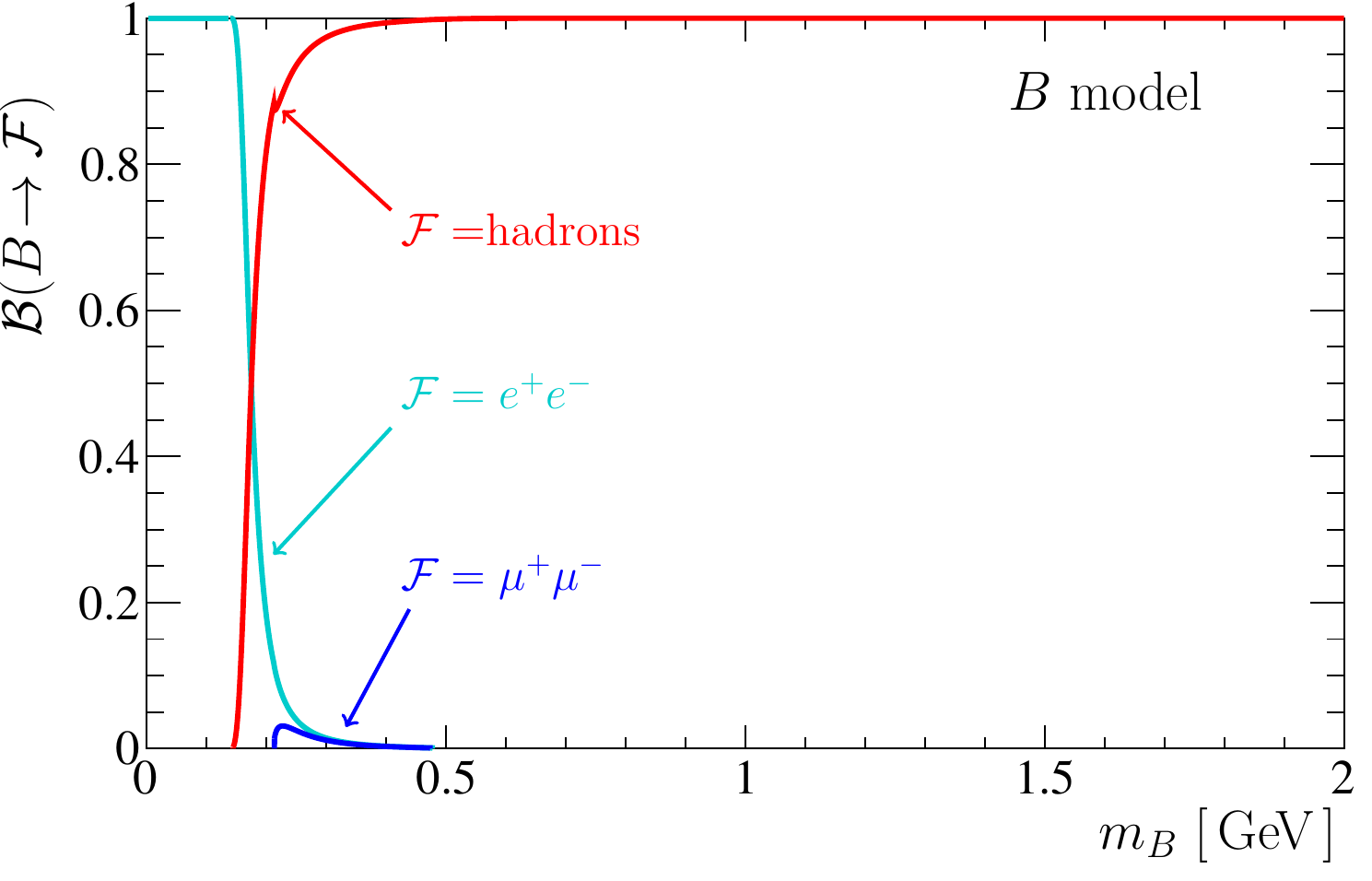}
  \includegraphics[width=0.49\textwidth]{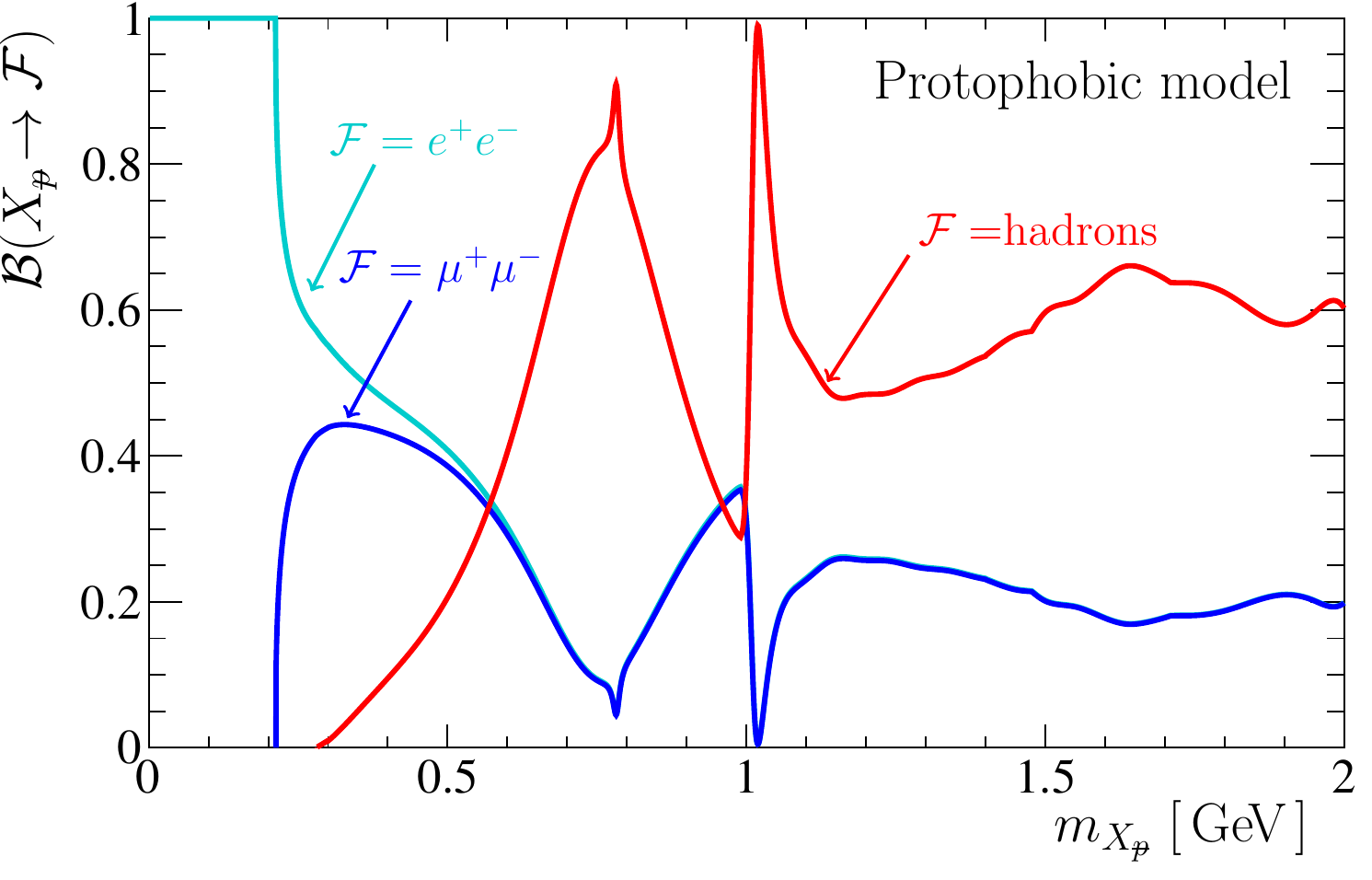}
  \includegraphics[width=0.49\textwidth]{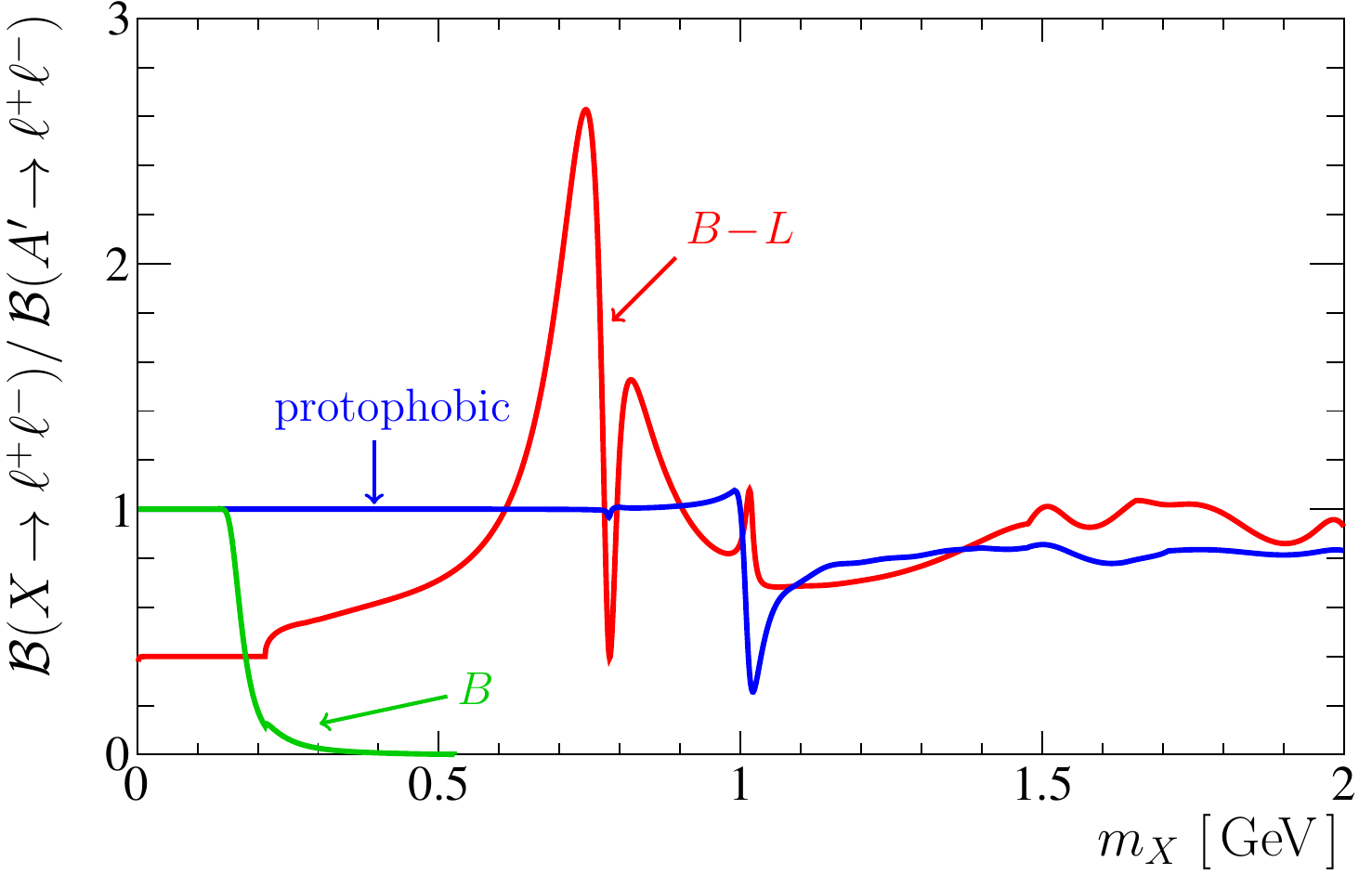}
  \caption{
  Decay branching fractions for the (top left)~\aprime, (top right)~\BL, (middle left)~$B$, and (middle right)~protophobic models. The branching fractions of the $B$ boson decaying into specific hadronic final states are shown in Fig.~\ref{fig:b_bfs}.
  (bottom) Ratio of the branching fractions to leptons for \BL, $B$, and the protophobic model relative to the \aprime.
  }
  \label{fig:bfs}
\end{figure*}

For the case where $\mathcal{B}(X\to\chi\bar{\chi})=0$,
it is straightforward to obtain all of the necessary $X$ decay branching fractions to SM final states, which are presented in Fig.~\ref{fig:bfs}.
These are determined using the couplings in Table~\ref{tab:models} and the results of Sec.~\ref{sec:xdec}, including the work in Appendix~\ref{app:xtohad}.
In addition, we provide the $\mathcal{B}_{B \to \mathcal{F}}$ values for all important decay modes of the $B$, including specific hadronic final states, in Fig.~\ref{fig:b_bfs} of Appendix~\ref{app:xtohad} as there are plans to use some of these final states in future searches (see, {\em e.g.}, Ref.~\cite{Fanelli:2016utb}).
The only hadronic final state used in any search considered here is $\aprime \to \pi^+\pi^-$, which was employed in the mass region near $m(\omega)$.
In this region, we take $\mathcal{B}_{X_{\pphob} \to \pi^+\pi^-} \approx \mathcal{B}_{\aprime \to \pi^+\pi^-}$, since the $\aprime$ and $X_{\pphob}$ both mix with the $\rho$ with equal strengths.
The decays $\BL \to \pi^+\pi^-$ and $B\to\pi^+\pi^-$ require isospin violation, making them difficult to calculate reliably.
One expects these branching fractions to be $\mathcal{O}(\%)$; however, we take them to be zero, since the only $\aprime \to \pi^+\pi^-$ search does not provide competitive sensitivity to $g_{\BL}$ or $g_B$.

\begin{figure*}[p]
  \centering
  \includegraphics[width=0.99\textwidth]{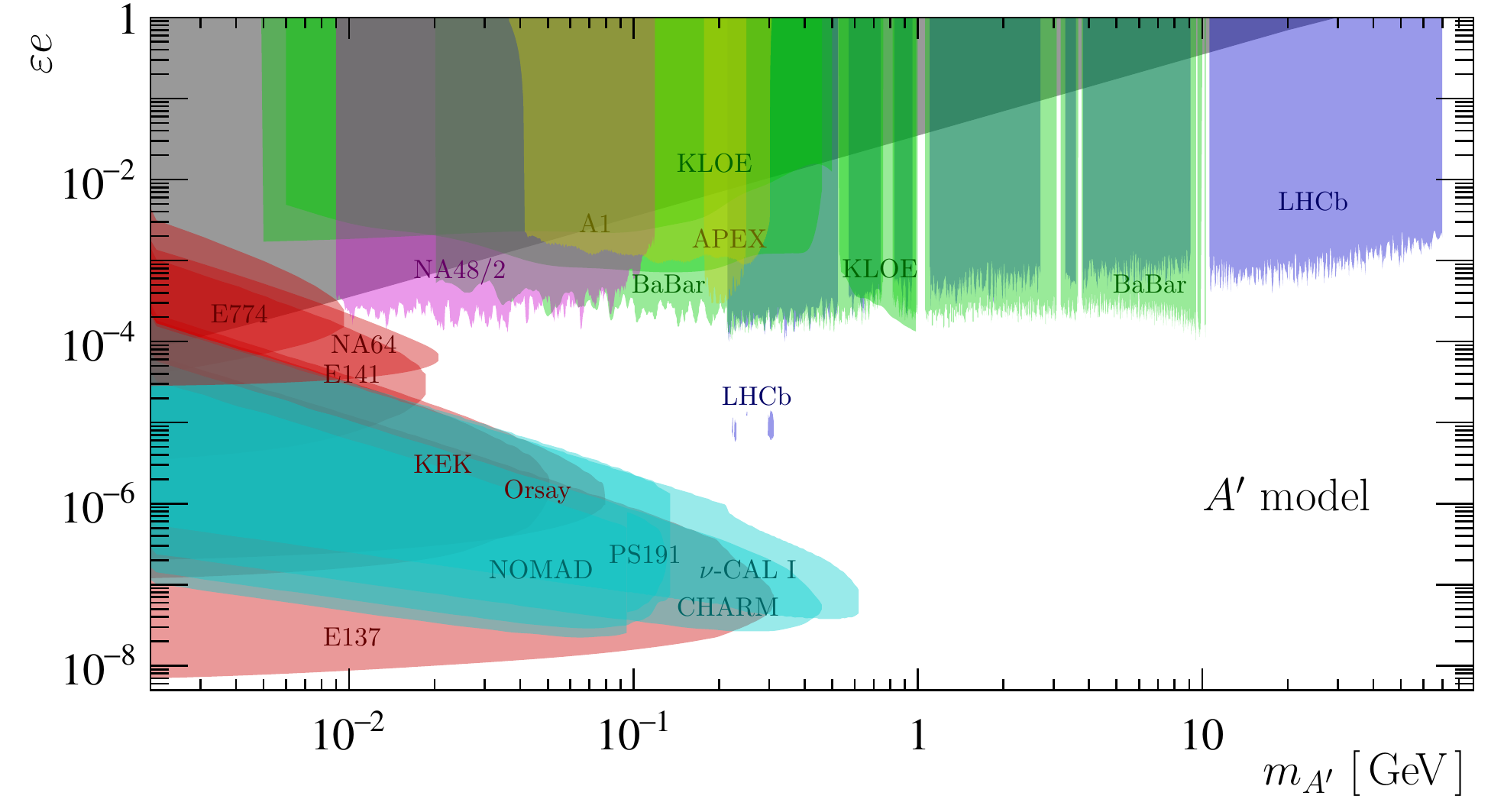}
  \caption{
  Constraints on visible \aprime decays considered in this study from
  (red) electron beam dumps,
  (cyan) proton beam dumps,
  (green) $e^+e^-$ colliders,
  (blue) $pp$ collisions,
  (magenta) meson decays,
  and (yellow) electron on fixed target experiments.
  The constraint derived from $(g-2)_e$ is shown in grey~\cite{Pospelov:2008zw,Endo:2012hp}.
  }
  \label{fig:results_aprime}
\end{figure*}

The searches for visible \aprime decays considered in our study are shown in Fig.~\ref{fig:results_aprime}.
We do not consider some searches that have inferior sensitivity to others that employed the same production and decay mechanisms.
The efficiency ratios are experiment dependent. Detailed discussion on these is provided in Appendix~\ref{app:exp}, see also Tables~\ref{tab:ExpPrompt} and \ref{tab:ExpBeamDump}.

\begin{figure*}[p]
  \centering
  \includegraphics[width=0.99\textwidth]{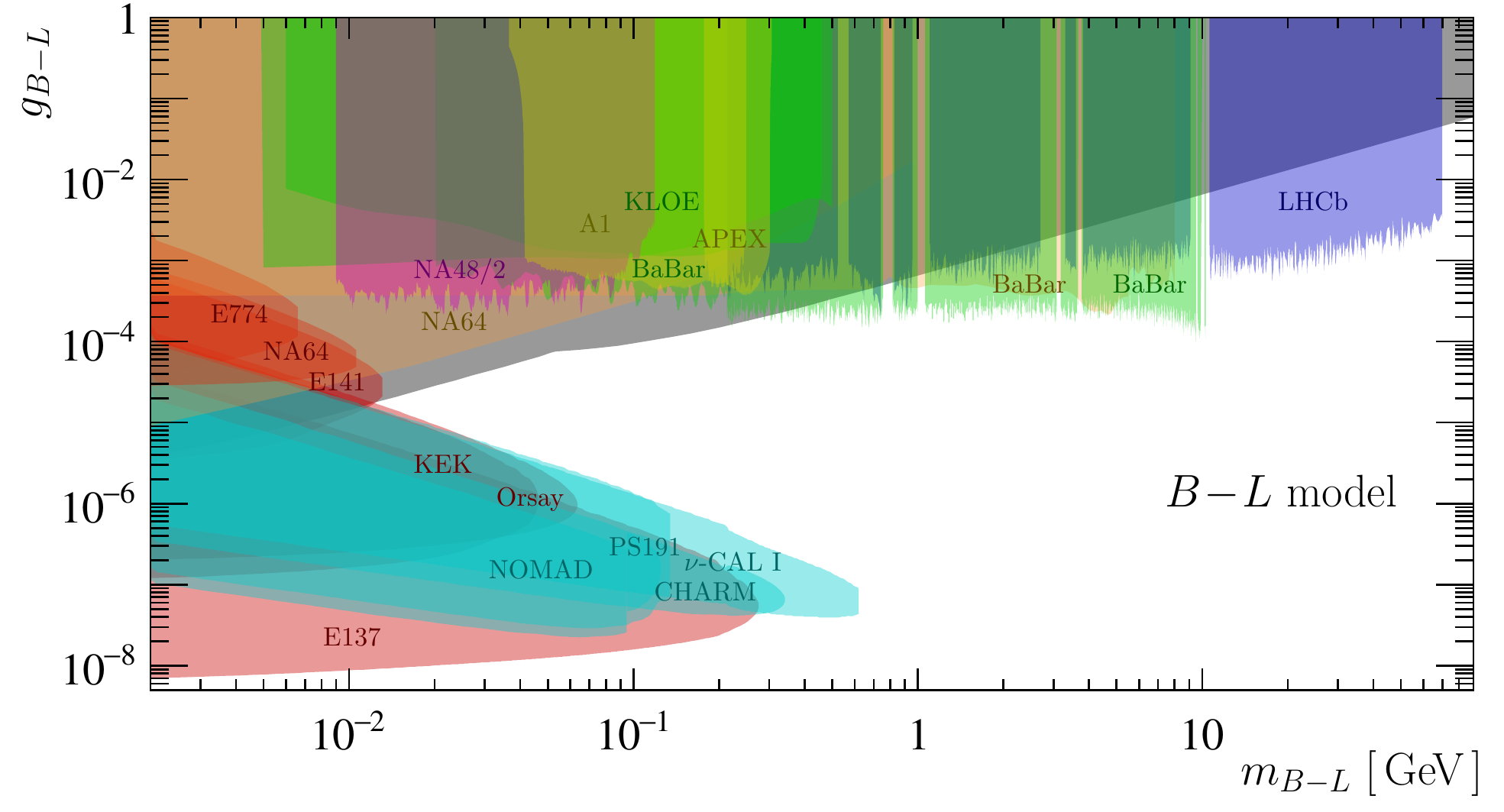}
  \caption{Constraints derived on \BL decays to SM final states using the same experimental color scheme as in Fig.~\ref{fig:results_aprime}.
  The (orange) invisible constraints also apply to \BL due to its coupling to neutrinos.
  The grey constraints are from Borexino~\cite{Harnik:2012ni,Bellini:2011rx}, Texono~\cite{Bauer:2018onh,Deniz:2009mu}, CHARM-II~\cite{Bauer:2018onh,Vilain:1993kd}, and from SPEAR, DORIS, and PETRA~\cite{Frugiuele:2016rii,Carlson:1986cu}.
  }
  \label{fig:results_bl}
\end{figure*}

\begin{figure*}[p]
  \centering
  \includegraphics[width=0.99\textwidth]{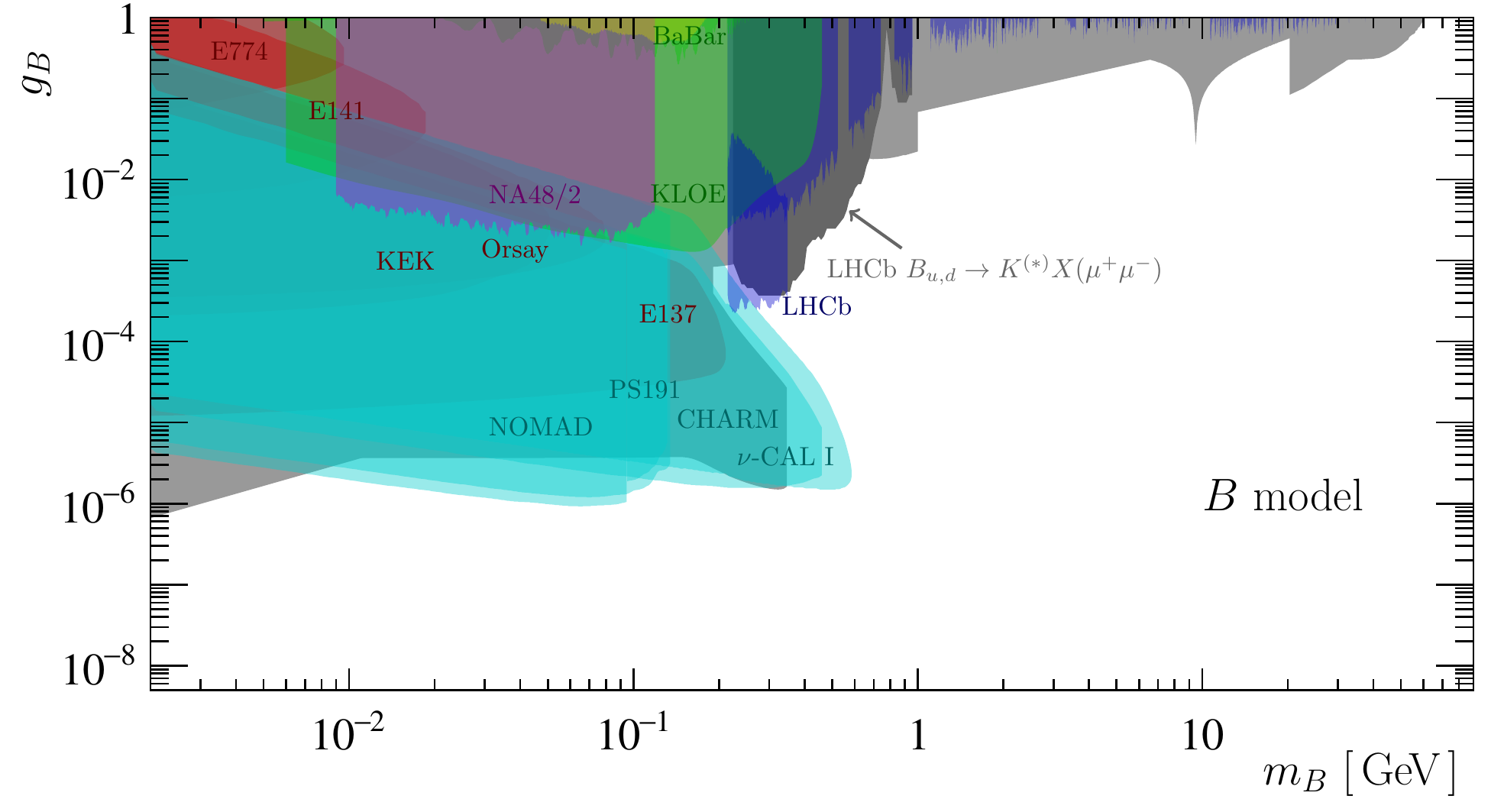}
\caption{Constraints derived on visible $B$ decays using the same experimental color scheme as in Fig.~\ref{fig:results_aprime}.  The grey constraints come from $\Upsilon$~\cite{Aranda:1998fr,Albrecht:1986ec} and $\eta$~\cite{Tulin:2014tya,Prakhov:2008zz} decays, from longitudinal-mode enhancements~\cite{Dror:2017ehi,Dror:2017nsg} in $B_{u,d}\to KX$~\cite{Grygier:2017tzo}, $K\to\pi X$~\cite{AlaviHarati:2003mr,Artamonov:2008qb}, and $Z\to X\gamma$~\cite{Acciarri:1997im,Abreu:1996pa} processes,
and from the lack of observed new anomaly-canceling fermions~\cite{Dobrescu:2014fca,Dror:2017ehi,Dror:2017nsg}.
The dark grey constraints, which are obtained in this work following Refs.~\cite{Dror:2017ehi,Dror:2017nsg}, are from the LHCb searches for $B_{u,d} \to K^{(*)}X$ with $X\to\mu^+\mu^-$~\cite{Aaij:2015tna,Aaij:2016qsm}.
}
  \label{fig:results_b}
\end{figure*}

\begin{figure*}[p]
  \centering
  \includegraphics[width=0.99\textwidth]{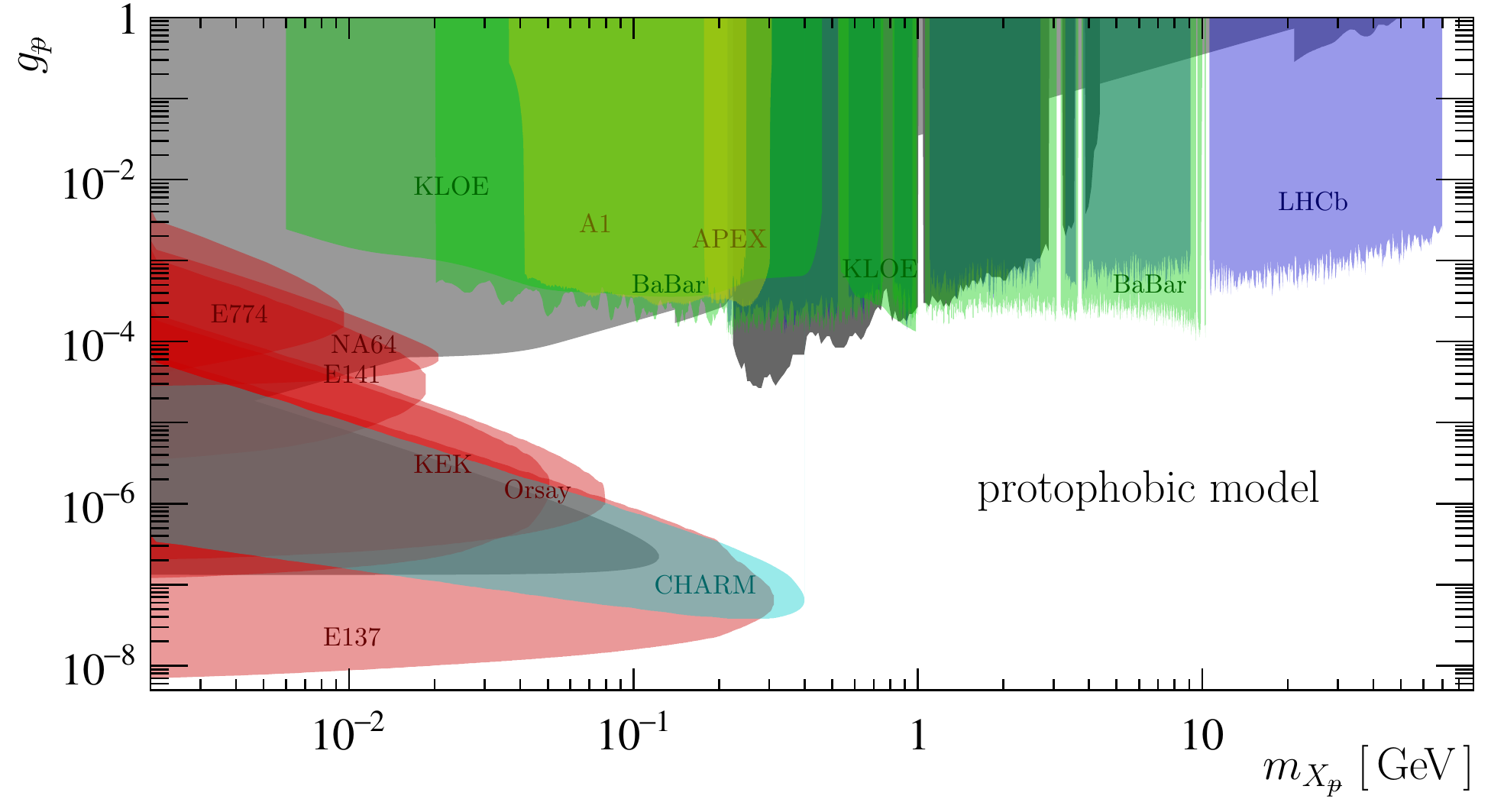}
  \caption{Constraints derived on visible protophobic decays using the same experimental color scheme as in Fig.~\ref{fig:results_aprime}.  The grey constraints are from the same processes as in Fig.~\ref{fig:results_b}, but recast to the protophobic model as part of this study.
   }
  \label{fig:results_p}
\end{figure*}

The \aprime results recast for \BL, $B$, and the protophobic model are shown in Figs.~\ref{fig:results_bl}--\ref{fig:results_p}.
Note that for the \BL model, which has nonzero couplings to SM neutrinos, searches for invisible dark photons also provide constraints  even for the $\mathcal{B}(\BL\to\chi\bar{\chi})=0$ case.
The recasted \aprime constraints on \BL, which are similar to the corresponding \aprime ones, are the strongest on this model in most of the coupling-mass region considered in Fig.~\ref{fig:results_bl}.
However, we note that recent constraints derived from neutrino experiments, where \BL exchange could compete with the SM neutral-current process, are currently the strongest available in a small region of $g_{\BL}$ values at small masses~\cite{Bauer:2018onh}.

For the $B$ model, the constraints bear little resemblance to those on the \aprime.
The lifetime of the $B$ is much larger than that of the \aprime for $g_B = \varepsilon e$ at low masses, due to the fact that the $B$ only couples to leptons via kinetic mixing.
One consequence of this is that the LHCb long-lived \aprime search~\cite{Aaij:2017rft} provides much better sensitivity to the $B$ boson than it does to the \aprime.
Since the $B$ couples to an anomalous SM current, additional strong constraints arise due to the enhanced production rates of the longitudinal $B$ mode as derived in Refs.~\cite{Dror:2017nsg,Dror:2017ehi}.
We have added to these the constraints from the LHCb searches for $B_{u,d} \to K^{(*)}X$ with $X\to\mu^+\mu^-$~\cite{Aaij:2015tna,Aaij:2016qsm}, which provide the strongest non-\aprime limits in the region $2m_{\mu} \lesssim m_B \lesssim 0.6\gev$.
Additional indirect constraints arise from the requirement of anomaly cancellation by new vector-like fermions, which have not yet been discovered~\cite{Dobrescu:2014fca}.
Under the assumption that the lack of discovery implies that such states do not exist, we apply these constraints following Refs.~\cite{Dror:2017nsg,Dror:2017ehi}, which are the strongest non-\aprime constraints in the mass region from about 1 to 5\gev.
The recasted \aprime constraints are the strongest on the $B$ at low masses, while the non-\aprime-search constraints are dominant for $m_B \gtrsim 0.4\gev$.

The constraints on the protophobic model are similar to those on the \aprime, except for the absence of the constraints based on production via proton bremsstrahlung and $\pi^0$ decays.
The protophobic current is also anomalous in the absence of additional fermions, which means that the constraints from Refs.~\cite{Dror:2017nsg,Dror:2017ehi} apply to this model as well; however, the coupling to the anomalous current is weaker by a factor of 4/9 due to the different fermionic couplings.\footnote{In the notation of Refs.~\cite{Dror:2017nsg,Dror:2017ehi}, the value of $\mathcal{A}_{XBB}$ is a factor of 4/9 smaller in the protophobic model than in the $B$ model.}
In addition, the sizable differences in the $X_{\pphob}$ lifetime and branching fractions lead to substantial differences in the constraints derived from the anomalous currents.
For example, the LHCb $B_{u,d} \to K^{(*)}X(\mu^+\mu^-)$ searches~\cite{Aaij:2015tna,Aaij:2016qsm} provide the strongest constraints in the region $2m_{\mu} \lesssim m_{X_{\pphob}} \lesssim 0.6\gev$ for the protophobic model.
That said, over most of the coupling-mass region explored thus far, the constraints obtained from \aprime searches are the most stringent.

\clearpage

\subsection{Decays to invisible dark-sector final states}
\label{sec:invsearches}

\begin{figure*}[t]
  \centering
  \includegraphics[width=0.99\textwidth]{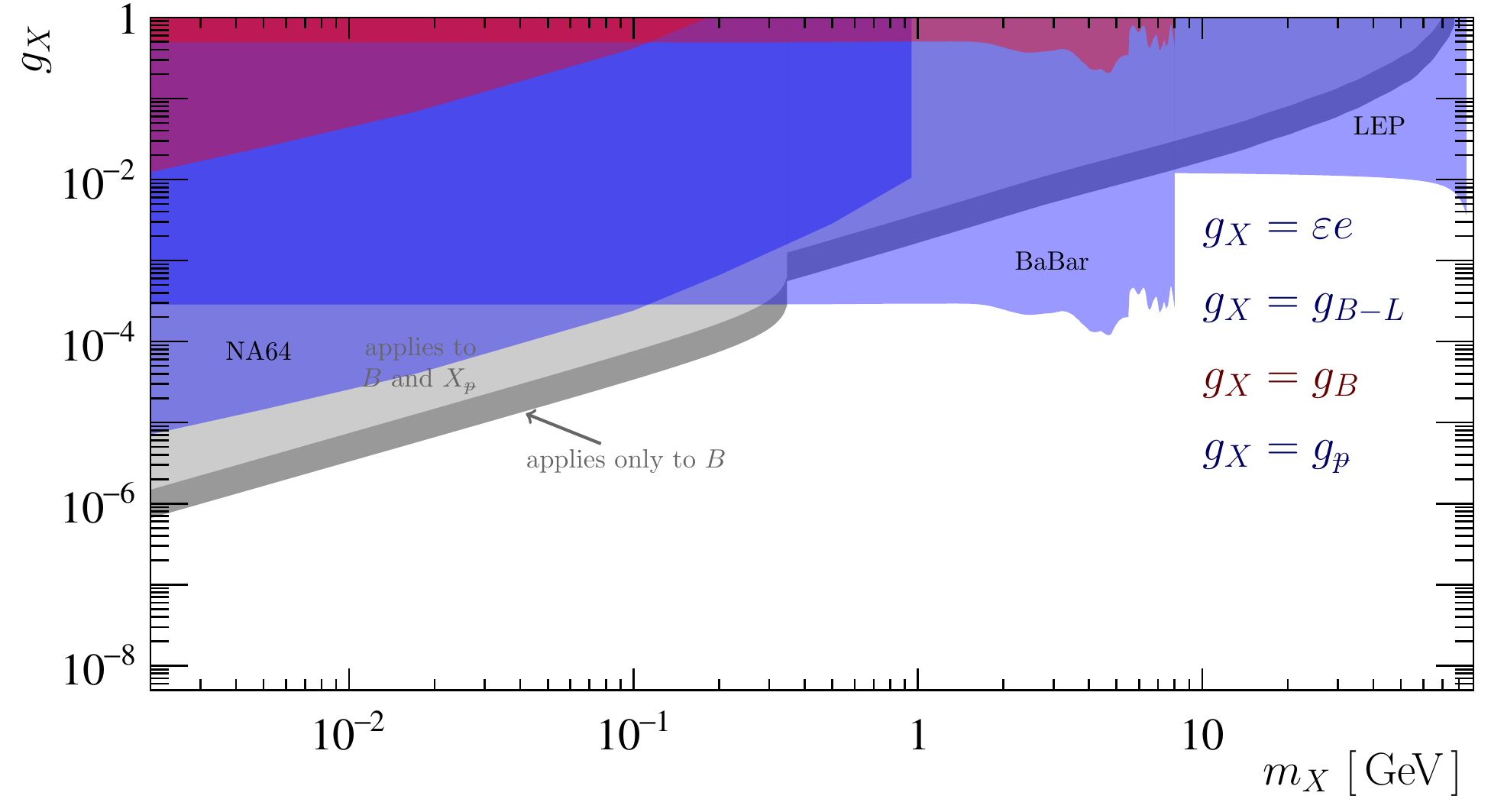}
  \caption{
  Constraints on all models assuming $\mathcal{B}(X\to\chi\bar{\chi}) \approx 1$. The grey constraints show the longitudinally enhanced results of Refs.~\cite{Dror:2017ehi,Dror:2017nsg} for $B$, also recast here for the protophobic model.
  }
  \label{fig:results_invisible}
\end{figure*}

For the case where $\mathcal{B}(X\to\chi\bar{\chi})\approx 1$, only the NA64~\cite{Banerjee:2017hhz}, BaBar~\cite{Lees:2017lec}, and LEP~\cite{Fox:2011fx} searches for dark photon decays to invisible final states are used in the recasting.
The results are shown in Fig.~\ref{fig:results_invisible}.
Additional constraints on the $B$ model, which couples to an anomalous SM current,
arise from $B_{u,d}\to KX$, $K\to\pi X$, and $Z\to X\gamma$ processes, as studied in Refs.~\cite{Dror:2017nsg,Dror:2017ehi}.
Recasting these results for the protophobic model, which also couples to an anomalous SM current, simply involves the scale factor of 4/9 discussed in the previous subsection.

\section{Summary}
\label{sec:sum}

In summary, we have developed a framework for recasting dark photon searches to obtain constraints on more generic models that contain a massive boson with vector couplings to the Standard Model fermions, which includes a data-driven method for determining hadronic decay rates.
We demonstrated our approach by deriving constraints on a vector that couples to the \BL current, a leptophobic $B$ boson that couples directly to baryon number and to leptons via $B$--$\gamma$ kinetic mixing, and on a vector that mediates a protophobic force.
This framework can easily be generalized to any massive boson with vector couplings to the Standard Model fermions (see, {\em e.g.}, Refs.~\cite{Fayet:1990wx,Campos:2017dgc}).
Of course, searches for dark photons can also provide sensitivity to non-vector particles~\cite{Kahn:2016vjr,Haisch:2016hzu}; however, recasting \aprime searches for scalars, {\em etc.}, does not lend itself to such a simple approach.
Finally, all information required to recast dark photon searches to any vector model, including software to perform any such recasting, is provided  at \git.

\section*{Acknowledgements}

We thank Jeff Dror, Stefania Gori, Maxim Pospelov, and Jesse Thaler for providing useful feedback.
In addition, we thank Jeff Dror for clarifying how the results in Refs.~\cite{Dror:2017ehi,Dror:2017nsg} were obtained,
Claudia Frugiuele and Elina Fuchs for providing the \BL constraints from Ref.~\cite{Frugiuele:2016rii},
Roni Harnik for giving additional information about Ref.~\cite{Fox:2011fx},
and Iftah Galon for useful discussion.
This work was supported by:
PI is supported by a Birmingham Fellowship;
YS is supported by the Office of High Energy Physics of the U.S.\ Department of Energy under grant contract number DE-SC00015476;
MW is supported by the U.S.\ National Science Foundation under contract number PHY-1607225;
WX is supported by the grant 669668-NEO-NAT-ERC-AdG-2014.

\appendix

\section{Additional VMD Details}
\label{app:vmd}

In this appendix, we provide additional details about the VMD calculations.
The most relevant $U(3)$ meson generators are
\begin{align}
& 	T_{\pi^0} = T_{\rho} = \frac{1}{2}{\rm diag}\{1,-1,0\} \, , \nonumber \\
& 	T_{\omega} = \frac{1}{2}{\rm diag}\{1,1,0\} \, , \nonumber\\
& 	T_{\phi} = \frac{1}{\sqrt{2}}{\rm diag}\{0,0,1\} \, ,  \\
&	T_{\eta} \approx \frac{1}{\sqrt{6}}{\rm diag}\{1,1,-1\} \, , \nonumber\\
& 	T_{\eta^{\prime}} \approx \frac{1}{2\sqrt{3}}{\rm diag}\{1,1,2\} \, , \nonumber\
\end{align}
using $\sin{\theta_{\rm mix}^{\eta,\eta^{\prime}}} \approx -1/3$ and $\cos{\theta_{\rm mix}^{\eta,\eta^{\prime}}} \approx 2\sqrt{2}/3$~\cite{Feldmann:1999uf}.
The VMD form factors are Breit-Wigner functions taken here to be
\begin{equation}
  	{\rm BW}_V(m) = \frac{m_V^2}{m_V^2 - m^2 - i m \Gamma_V(m)} \, ,
\end{equation}
where the mass-dependent widths, which account for changes in the kinematic factors in both the decay amplitudes and phase space collectively denoted by $\mathcal{K}_{\mathcal{F}}(m)$ for the decay $V\to\mathcal{F}$ (see, {\em e.g.}, Refs.~\cite{Achasov:2003ir,Achasov:2002ud,Lees:2013gzt} for these kinematic factors), are
\begin{equation}
 	\Gamma_V(m) = \sum_{\mathcal{F}} \mathcal{B}_{V\to\mathcal{F}} \Gamma_V(m_V) \frac{\mathcal{K}_{\mathcal{F}}(m)}{\mathcal{K}_{\mathcal{F}}(m_V)} \, .
\end{equation}
The following final states are considered for $\Gamma_V(m)$:
$\pi^+\pi^-$ for the $\rho$\,;
$\pi^+\pi^-\pi^0$, $\pi^0\gamma$, and $\pi^+\pi^-$ for the $\omega$\,;
and $K^+K^-$, $K_SK_L$, $\pi^+\pi^-\pi^0$, and $\eta\gamma$ for the $\phi$\,.
Finally, for both gauged \BL and $B$, the quark couplings are universal and given by
\begin{equation}
Q_{\BL} = Q_B = \frac{1}{3}{\rm diag}\{1,1,1\},
\end{equation}
while for the protophobic force the  quark-coupling matrix is
\begin{equation}
  Q_{\pphob} = \frac{1}{3}{\rm diag}\{-1,2,2\}.
\end{equation}
The most relevant decay rates for producing these bosons are listed in Table~\ref{tab:meson_decays}.

\section{$X\to$\,hadrons}
\label{app:xtohad}

To obtain reliable predictions of $\Gamma_{X\to{\rm hadrons}}$ for low masses, we have developed a data-driven approach based on measured $e^+e^-\to\mathcal{F}$ cross sections.
As stated above, we first normalize each of the most important low-mass hadronic $e^+e^-\to\cF$ cross sections to that of $e^+e^- \to \mu^+\mu^-$ according to Eq.~\eqref{eq:Rfit}.
The $\cA_{\cF}^V$ amplitudes in Eq.~\eqref{eq:Rfit2} are given by
\begin{equation}
  	\cA_{\cF}^V(m)
\!=\! 	\frac{\Gamma_V}{m_V} {\rm BW}_V(m) \sqrt{\frac{\cB_{V\to e^+e^-}\cB_{V\to\cF}\cK_{\mathcal{F}}(m)}{\cK_{\cF}(m_V)}}  .
\end{equation}
We then fit the $e^+e^-\to\cF$ cross-section data for the most important hadronic final states, and use these results to
decompose $e^+e^-\!\!\to$\,hadrons into $\rho$-like, $\omega$-like, and $\phi$-like contributions (see Fig.~\ref{fig:rmu_data2}) defined as:
\begin{itemize}
  \item The dashed $\gamma$-like line shows the sum of all final states considered here, including $\pi^0\gamma$, which overshoots\,(undershoots) the PDG $\cR_{\mu}$ data for ${m \lesssim 1.5\gev}$\,($m \gtrsim 1.5\gev$).
  The PDG result was produced in 2003, and it does not include any of the high-precision data used in our study.\footnote{Ref.~\cite{TheBaBar:2017vzo} shows a comparison of the recent BaBar $\pi^+\pi^-\pi^0\pi^0$ data to the older data used to make the PDG average, where one can see that the dip in the PDG data at $m \approx 1.45\gev$ is most likely an experimental artifact that arose due to a confluence of experimental thresholds.}
  We take the total  $e^+e^-\!\!\to$\,hadrons---the solid $\gamma$-like line in Fig.~\ref{fig:rmu_data2}---to be our sum below $1.48\gev$ and the PDG version otherwise, since at higher masses decay modes not included in our study are expected to be important.
  \item The dashed $\omega$-like curve includes the $\omega\to\pi^0\gamma$ contribution, along with the model used to fit the $\pi^+\pi^-\pi^0$ data but with the $\phi$ amplitude removed. Interference between the $\phi$ and $f_{\pi^+\pi^-\pi^0}(m)$ terms causes the large visible dip near 1.05\gev, which is far from any $\omega^*$ poles justifying the use of a real $f_{\pi^+\pi^-\pi^0}(m)$ function.
  The LO perturbative value of $\mathcal{R}_{\mu}^{\omega}$ is 1/6. The $\omega$-like curve overshoots this slightly near 1.6\gev, which is not unexpected given that there are several $\omega^*$ poles nearby, then falls rapidly at higher masses. We assume that this fall off is due to additional (neglected) final states becoming important, and augment the $\omega$-like contribution (solid curve) to take on the LO perturbative value for $m \gtrsim 1.6\gev$.
  \item The dashed $\phi$-like curve includes the $KK$ and $\left[KK\pi\right]_{I=0}$ contributions, along with $\phi \to \pi^+\pi^-\pi^0$.
  The LO perturbative value of $\mathcal{R}_{\mu}^{\phi}$ is 1/3.
  Similarly to the $\omega$-like curve, the $\phi$-like curve is expected to overshoot the LO perturbative value near the $\phi(1680)$, and the fact that it falls off at higher masses is assumed to be due to neglected final states. We augment the $\phi$-like contribution (solid curve) to take on the LO perturbative value for $m \gtrsim 1.7\gev$.
  \item Finally, the $\rho$-like contribution is assumed to be entirely described by the $\pi^+\pi^-$ and $4\pi$ data for $m < 1.1\gev$, and is defined as the (solid) $\gamma$-like contribution with the (solid) $\omega$-like and $\phi$-like curves subtracted for $m > 1.1\gev$. The resulting $\rho$-like curve is within 10\% of its LO perturbative value of 3/2 for $m \gtrsim 1.8\gev$.\footnote{This approach attributes all of the $\rho$--$\omega$ mixing in the $\pi^+\pi^-$ final state to the $\rho$-like current. While one could certainly question the validity of this choice, the level at which isospin violation occurs in vector mesons is small compared to the overall precision of the VMD calculations for production rates; therefore, it is acceptable to neglect this complication when recasting the dark photon results.}
\end{itemize}
We can further justify the use of the LO perturbative values at higher masses by the fact that $\cR_{\mu}$ itself is within 20\% of its LO perturbative value of 2 for $m \gtrsim 1.5\gev$.

\begin{figure*}[t]
  \centering
  \includegraphics[width=0.99\textwidth]{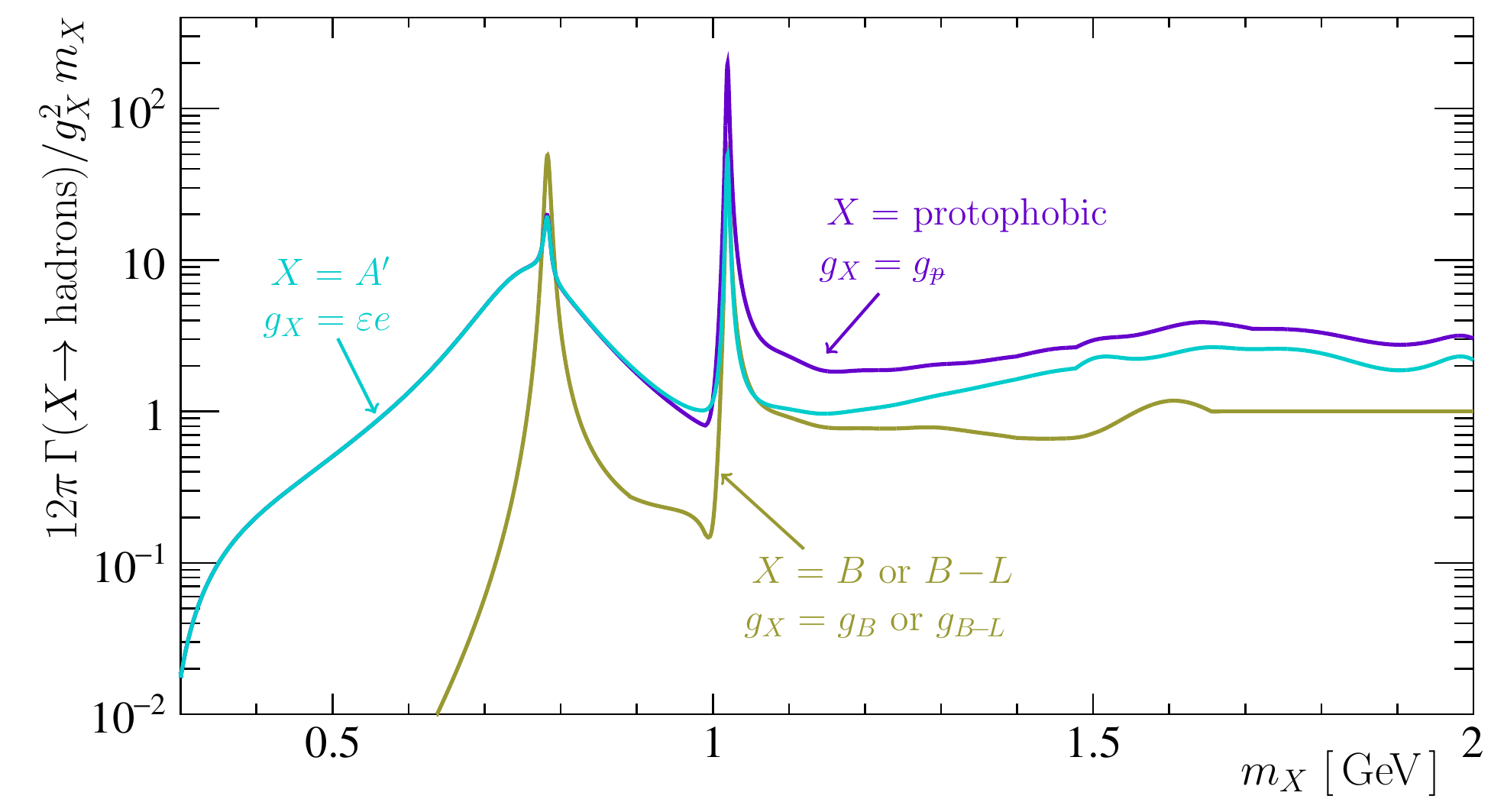}
  \caption{
    Hadronic decay width for: a dark photon, where $g_X = \varepsilon e$;  a gauged \BL or $B$ boson, where $g_X = g_B$ or $g_{\BL}$;
  and a protophobic boson, where  $g_X = g_{\pphob}$.
  }
  \label{fig:gamma_had}
\end{figure*}

\begin{figure*}[t]
  \centering
  \includegraphics[width=0.99\textwidth]{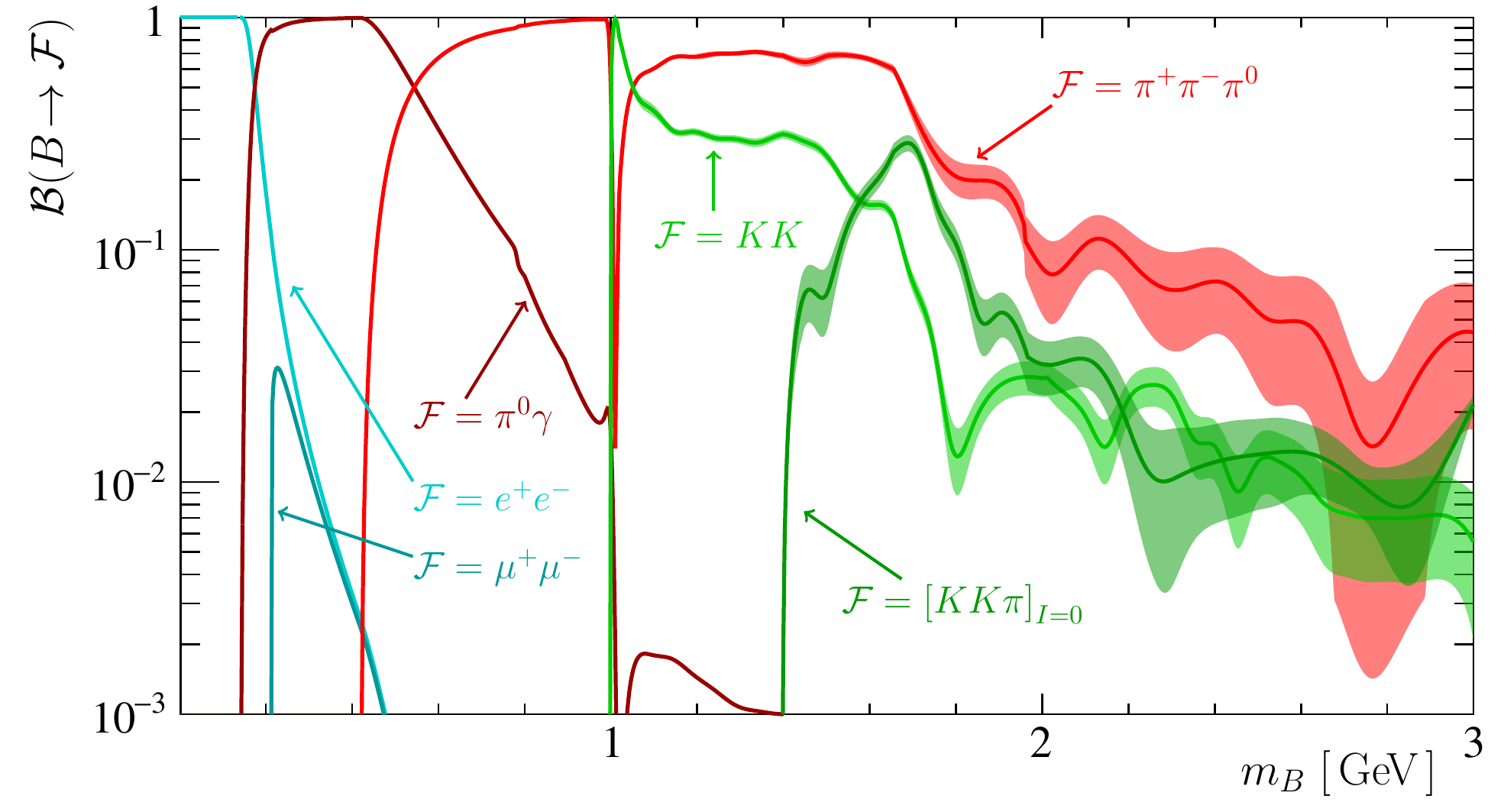}
  \caption{
  Decay branching fractions for the $B$ boson assuming a kinetic-mixing parameter $\varepsilon = e g_B / (4\pi)^2$. The error bands shown for the final states $\pi^+\pi^-\pi^0$, $KK\equiv K^+K^- + K_SK_L$,
  and $\left[KK\pi\right]_{I=0}$ ({\em i.e.}\ the isoscalar component of the $KK\pi$ final state) are due to the limited experimental knowledge of the $e^+e^-\!\!\to\mathcal{F}$ cross sections. In addition, the use of VMD and $U(3)$ symmetry introduces roughly a 20\% uncertainty on all hadronic decay rates.}
  \label{fig:b_bfs}
\end{figure*}

Using these $\rho$-like, $\omega$-like, and $\phi$-like models, we can estimate $\Gamma_{X \to {\rm hadrons}}$ for any $X$ model using Eq.~\eqref{eq:GammaXhad}.
Figure~\ref{fig:gamma_had} shows $\Gamma_{X \to {\rm hadrons}}$ for a dark photon, along with for the \BL, $B$, and protophobic models.
By construction, our approach gives the canonical $\Gamma_{\aprime \to {\rm hadrons}}$ result for the dark photon model.\footnote{With the caveat of using an updated $\mathcal{R}_{\mu}$ for $m \lesssim 1.6\gev$.}
Since \BL and $B$ do not mix with the $\rho$, their hadronic decay rates are substantially lower, especially at lower masses.
Note that the $\phi$--$\omega$ interference dip is below the $\phi$ peak for these models, since the relative sign between the $\omega$ and $\phi$ amplitudes is positive here versus negative for the \aprime model.
The protophobic model has a similar hadronic decay width to the \aprime below the $\phi$; however, at larger masses its width is larger due to its larger $s$-quark coupling.
Finally, we also provide the $\mathcal{B}_{B \to \mathcal{F}}$ values for all important decay modes of the $B$, including specific hadronic final states, in Fig.~\ref{fig:b_bfs} as there are plans to use some of these final states in future searches~\cite{Fanelli:2016utb}.

\clearpage

\section{Experiments}
\label{app:exp}

This section contains all of the experiment-specific information, which is summarized Tables~\ref{tab:ExpPrompt} and \ref{tab:ExpBeamDump}.

\begin{table*}
  \begin{center}
    \caption{\label{tab:ExpPrompt} Summary of experiments that searched for prompt or invisible \aprime decays. See Appendix~\ref{app:exp.lhcb} for discussion on the LHCb search for long-lived \aprime decays.}
    {\renewcommand{\arraystretch}{2}
      \begin{tabular}{c|c|c|c}
        experiment & production  & final state & efficiency ratio $\displaystyle\epsilon(\tau_X)/\epsilon(\tau_{\aprime})$  \\
        \hline
        BaBar~\cite{Lees:2014xha}  & $e^+e^-\to X\gamma$ & $e^+e^-$, $\mu^+\mu^-$ & 1\\
        NA48/2~\cite{Batley:2015lha} & $\pi^0\to X\gamma$ & $e^+e^-$ & Eq.~\eqref{eq:prompt_eff} with $\tilde{t} = [1\,{\rm m}] / (c\,\gamma)$,\vspace{-0.2in}\\
        {} & {} & {} &  where $\gamma = 50\gev/ m_X$\\
        A1~\cite{Merkel:2014avp}, APEX~\cite{Abrahamyan:2011gv} & $eZ\to eZ X$ & $e^+e^-$ & 1 \\
        KLOE~\cite{Archilli:2011zc} & $\phi\to X\eta$ & $e^+e^-$ & Eq.~\eqref{eq:prompt_eff} with $\tilde{t} = [8\,{\rm cm}] / (c\,\gamma)$, \vspace{-0.07in}\\
        {} & {} & {} & where $\gamma = \displaystyle\frac{m_{\phi}^2 + m_X^2 - m_{\eta}^2}{2m_{\phi} m_X}$ \\
        KLOE~\cite{KLOE:2016lwm} & $e^+e^-\to X\gamma$ & $\pi^+\pi^-$ & 1   \\
        KLOE~\cite{Anastasi:2015qla} & $e^+e^-\to X\gamma$ & $e^+e^-$ & 1   \\
        LHCb~\cite{Aaij:2017rft} & inclusive $pp\to X$ & $\mu^+\mu^-$ &   Eq.~\eqref{eq:prompt_eff} with  \vspace{-0.2in}\\
        {} & see Fig.~\ref{fig:lhcb_fracs} & &  $\tilde{t} \approx \{[4\mev]/(m_X\!-\!2m_{\mu})\!+\!0.1\}\ps$ \\
        \hline
        BaBar~\cite{Lees:2017lec}  & $e^+e^-\to X\gamma$ & invisible & 1\\
        NA64~\cite{Banerjee:2017hhz} & $eZ\to eZX$ & invisible & 1 \\
        LEP~\cite{Abdallah:2003np,Abdallah:2008aa} & $e^+e^-\to X\gamma$~\cite{Fox:2011fx} & invisible & 1
      \end{tabular}}
  \end{center}
\end{table*}

\begin{table*}
  \begin{center}
    \caption{\label{tab:ExpBeamDump} Summary of beam dump experiments that searched for long-lived dark photons.}
    {\renewcommand{\arraystretch}{3}
      \begin{tabular}{c|c|c|c}
        experiment & production  & final state & $L_{\rm dec}/L_{\rm sh}$  \\
        \hline
        E141~\cite{Riordan:1987aw}, E137~\cite{Bjorken:1988as}, E774~\cite{Bross:1989mp}, & $eZ\to eZX$~\cite{Bjorken:2009mm,Andreas:2012mt} & $e^+e^-$ & $\displaystyle\frac{35}{0.12}$, $\displaystyle\frac{204}{179}$, $\displaystyle\frac{2}{0.3}$,  \vspace{-0.2in}\\ 
         KEK~\cite{Konaka:1986cb}, Orsay~\cite{Davier:1989wz} & & & $\displaystyle\frac{2.2}{2.4}$,  $\displaystyle\frac{2}{1}$ \vspace{-0.2in}\\
         NA64~\cite{Banerjee:2018vgk} & & & $\approx 4$ \\
        $\nu$-CAL~I~\cite{Blumlein:1990ay,Blumlein:1991xh} & $\pi^0\to X\gamma$~\cite{Blumlein:2011mv},  & $e^+e^-$ &   $\displaystyle\frac{23}{64}$  \vspace{-0.3in} \\ 
        {} & $pZ\to pZX$~\cite{Blumlein:2013cua} & {} & \\
        CHARM~\cite{Bergsma:1985qz} & $\eta^{(\prime)}\to X\gamma$~\cite{Gninenko:2012eq} & $e^+e^-$ & $\displaystyle\frac{10}{480}$  \\
        NOMAD~\cite{Astier:2001ck}, PS191~\cite{Bernardi:1985ny} & $\pi^0\to X\gamma$~\cite{Gninenko:2011uv} & $e^+e^-$ &  $\displaystyle\frac{7.5}{835}$, $\displaystyle\frac{7}{128}$  
      \end{tabular}}
  \end{center}
\end{table*}

\subsection{BaBar}

The BaBar collaboration published strong constraints on both visible~\cite{Lees:2014xha} and invisible~\cite{Lees:2017lec} \aprime decays.
Their visible search used both $\aprime \to e^+e^-$ and $\aprime \to \mu^+\mu^-$, and required that the leptons were consistent with originating from the beam interaction region.
Even for models where $\Gamma_{X \to {\rm hadrons}}=0$, the BaBar visible search is not sensitive to $X$ bosons with lifetimes large enough to qualify as non-prompt; therefore, the efficiency ratio $\epsilon(\tau_X)/\epsilon(\tau_{\aprime})$ is unity.
The BaBar visible results combine the two $\ell^+\ell^-$ final states.
To recast this search for the case where $x_e \neq x_{\mu}$, the individual limits provided in the Supplemental Material of Ref.~\cite{Lees:2014xha} should be used.
In this work, we only consider models with $x_e = x_{\mu}$, where the recasted constraints are obtained by solving
\begin{equation}
  	(g_X x_e)^2
= 	(\varepsilon e)^2 \frac{\mathcal{B}(\aprime \to \ell^+\ell^-)}{\mathcal{B}(X\to \ell^+\ell^-)} \, .
\end{equation}
For the invisible search, the assumption is again that the efficiency ratio is unity and the branching-fraction ratio above is replaced by the equivalent ratio into invisible final states.

\subsection{NA48/2}

The NA48/2 experiment searched for $\pi^0\to\aprime\gamma$ followed by prompt $\aprime \to e^+e^-$ decays~\cite{Batley:2015lha}.
The prompt requirement maintains high efficiency until the flight distance reaches about 1\,m.
The maximum $\gamma$ factors are about $50\gev/m_X$.
We take the prompt-criteria efficiency to be given by Eq.~\eqref{eq:prompt_eff} with
$\tilde{t} = [1\,{\rm m}] / (c\,\gamma)$ and $\gamma = 50\gev/ m_X$,
which is unity for the \aprime.
This efficiency factor, however, is important for a leptophobic boson, since the production utilizes the quark couplings whereas the decay must go to $e^+e^-$, which is suppressed as it arises due to kinetic mixing.
Recasting these limits for an $X$ boson is done using Eq.~\eqref{eq:PVG}, see also Table~\ref{tab:meson_decays}, with the appropriate mass- and model-dependent values of $\mathcal{B}(X\to e^+e^-)$ and $\mathcal{B}(\aprime \to e^+e^-)$.

\subsection{Electron Bremsstrahlung}

The A1~\cite{Merkel:2014avp} and APEX~\cite{Abrahamyan:2011gv} experiments provide the best electron bremsstrahlung constraints on promptly decaying dark photons.
The decay $\aprime \to e^+e^-$ was searched for by both experiments, and the recasting is done using
\begin{equation}
  	(g_X x_e)^2 = (\varepsilon e)^2 \frac{\mathcal{B}(\aprime \to \ell^+\ell^-)}{\mathcal{B}(X\to \ell^+\ell^-)}.
\end{equation}
Neither experiment provides detailed information about prompt-like requirements; however, since the same coupling is used to produce and decay the boson, it is safe to simply take the efficiency ratio with the \aprime to be unity for all $X$ models.
Additionally, the NA64 experiment at CERN used 100\gev electrons incident on an active target to search for invisible \aprime decays~\cite{Banerjee:2016tad,Banerjee:2017hhz}.
For this search, the assumption is that the efficiency ratio is unity and the branching-fraction ratio above is replaced by the equivalent ratio into invisible final states.

\subsection{KLOE}

The KLOE experiment searched for $\phi \to \aprime \eta$ followed by a prompt $\aprime \to e^+e^-$ decay~\cite{Archilli:2011zc}.
Our interpretation of the prompt criteria is that good efficiency should be maintained provided that the flight distance is $\lesssim 8$\,cm.
The $\gamma$ factors here are  $(m_{\phi}^2 + m^2 - m_{\eta}^2)/(2m_{\phi} m)$, which are $\mathcal{O}(1\text{--}10)$ in the mass range where KLOE has good sensitivity.
We take the prompt-criteria efficiency to be given by Eq.~\eqref{eq:prompt_eff} with $\tilde{t} = [8\,{\rm cm}] / (c\,\gamma)$, which is unity for the \aprime.
Recasting these limits uses Eq.~\eqref{eq:VVP} but taking the sum ${\mathcal{B}(\aprime \to e^+e^-)+ \mathcal{B}(\aprime \to \mu^+\mu^-)=1}$, which was assumed by KLOE, along with the $\tau$-dependent efficiency factor for the $X$.
{\em N.b.}, since this search involves an $X$ produced via quark couplings and decaying via leptonic couplings, the $\tau$-dependent efficiency factor can be important despite being $\approx 100\%$ efficient for the \aprime.

KLOE also searched for $e^+e^-\!\! \to \aprime \gamma$ using the ${\aprime \to \pi^+\pi^-}$ decay~\cite{KLOE:2016lwm}.
The pions were required to have their points of closest approach to the beam line within a cylindrical volume of radius 8\,cm and length 15\,cm.
The $\gamma$ factors in this search are $(m_{\phi}^2 + m^2)/(2m_{\phi} m)\lesssim 1.2$, which means that inefficiency due to the prompt criteria should only arise for ${c \tau_X \gtrsim \mathcal{O}(10\,{\rm cm})}$, which is not the case for any of the models studied in this work.
Recasting these results is done using
\begin{equation}
  (g_X x_e)^2 = (\varepsilon e)^2 \frac{\mathcal{B}(\aprime \to \pi^+\pi^-)}{\mathcal{B}(X\to \pi^+\pi^-)}.
\end{equation}
This search is useful because it fills in the gap near the $\omega$ peak in the \aprime constraints.
Finally, KLOE performed a similar search looking for prompt ${\aprime \to e^+e^-}$ decay~\cite{Anastasi:2015qla}.
In this search, the cylindrical decay volume used had a radius of 1\,cm and a length of 12\,cm, which is sufficiently large that it does not induce any lifetime-based inefficiencies in any of the models studied here.

\subsection{LHCb}
\label{app:exp.lhcb}

\begin{figure*}[t]
  \centering
  \includegraphics[width=0.99\textwidth]{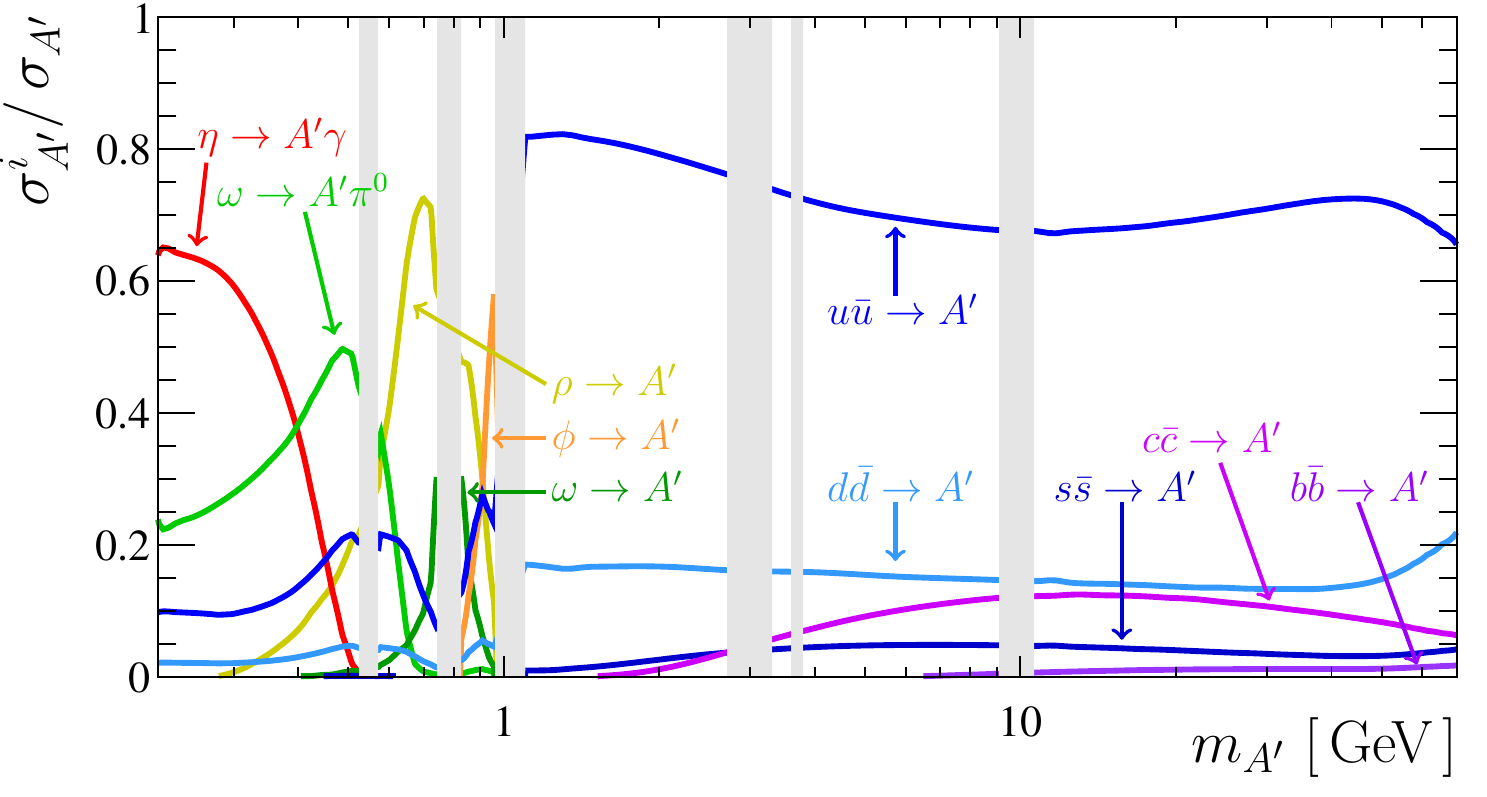}
  \caption{
  Dark-photon production fractions for the most important processes at LHCb in the fiducial region of Ref.~\cite{Aaij:2017rft}.
  }
  \label{fig:lhcb_fracs}
\end{figure*}

An inclusive search for dark photons using the $\aprime \to \mu^+ \mu^-$ decay was performed by the LHCb experiment~\cite{Aaij:2017rft}.
Both prompt and long-lived limits were published, where the latter provide $r^{\rm ul}_{\rm ex}$ as a function of $\ma$ and $\varepsilon^2$.
Consequently, the only information needed to recast the LHCb results is the relative fraction of each \aprime production mechanism as a function of $\ma$, as given in Fig.~\ref{fig:lhcb_fracs}.
We determine these ratios by fitting the inclusive $\mu^+ \mu^-$ background-subtracted mass spectrum published by LHCb, using Monte Carlo signal templates generated using \textsc{Pythia 8}~\cite{Sjostrand:2014zea}.
Only templates for the following predominant production mechanisms are considered in the fit:
$\eta \to \mu^+ \mu^- \gamma$, $\eta \to \mu^+ \mu^-$, $\omega \to \mu^+ \mu^- \pi^0$, $\omega \to \mu^+ \mu^-$, $\rho \to \mu^+ \mu^-$, $\phi \to \mu^+ \mu^-$ and Drell-Yan.

All of the fiducial requirements applied in the LHCb analysis are applied to the Monte Carlo dimuons when obtaining the templates.
The nominal fractions are obtained using the cross-sections predicted with \textsc{Pythia 8},
combined with the relevant measured branching fractions~\cite{Patrignani:2016xqp}.
Each template is smeared to account for the LHCb mass resolution.
The $\eta \to \mu^+ \mu^- \gamma$ and $\omega \to \mu^+ \mu^- \pi^0$ mass shapes from \textsc{Pythia 8} are generated using a generic VMD-based Dalitz decay,
and so these two templates are corrected using the mass shapes obtained from an NA60 analysis~\cite{Arnaldi:2016pzu}.
Similarly, the $\rho \to \mu^+ \mu^-$ mass shape is also corrected using the same NA60 analysis, while the $\omega \to \mu^+ \mu^-$ and $\phi \to \mu^+ \mu^-$ mass shapes are corrected using the results of Ref.~\cite{Achasov:2003ir}.

The LHCb mass spectrum is fitted by allowing the fraction for each template to vary within $0$ to $10$ times its nominal value,
where the total Drell-Yan production is considered as a single template.
The same isolation criterion applied in the LHCb analysis above the $\phi$ mass was also applied to the Monte Carlo dimuons when building the templates.
However, the isolation quantity is not expected to match exactly between the Monte Carlo and data due to reconstruction effects.
Therefore, the efficiency of this isolation requirement is also allowed to vary in the fit, resulting in a total of $8$ free parameters.
A validation of the fit is that the ratios of the two $\eta$ and the two $\phi$ channels match their respective known values within uncertainties.

From the Supplemental Material to Ref.~\cite{Aaij:2017rft}, one can see that the LHCb prompt-selectrion criteria are the same as those we proposed in Ref.~\cite{Ilten:2016tkc};
therefore, we use our simulation samples from that study and find that the efficiency is well approximated by Eq.~\eqref{eq:prompt_eff} with $\tilde{t} \approx \{[4\mev]/(m_X-2m_{\mu})+0.1\}\ps$.
As discussed above, since LHCb published $r^{\rm ul}_{\rm ex}$ as a function of $\ma$ and $\varepsilon^2$, recasting the long-lived \aprime search can be done using Eq.~\eqref{eq:lhcb_displ_recast}.
For $\tau_X$ values that fall outside of the range where LHCb provided results, the efficiency ratios are taken to be
\begin{equation}
  \frac{\epsilon(\tau_X)}{\epsilon(\tau_{\aprime}^{\rm min,max})} \approx
  \begin{cases}
    e^{1-\left(\tau_{\aprime}^{\rm min}/\tau_X\right)} & \text {for}\, \tau_X < \tau_{\aprime}^{\rm min}, \\
    \frac{1-e^{-\tau_{\aprime}^{\rm max}/\tau_X}}{1-e^{-1}} & \text {for}\, \tau_X > \tau_{\aprime}^{\rm max}, \\
  \end{cases}
\end{equation}
which correspond to a long-lived selection efficiency of zero for decay times less than the minimum reported by LHCb (justified by the efficiency figure provided in the Supplemental Material of Ref.~\cite{Aaij:2017rft})
and to a maximum decay time that results in the muons being reconstructed by the first LHCb tracking system being less than the maximum $\tau_{\aprime}$ reported by LHCb (confirmed to be a good approximation by our simulation from Ref.~\cite{Ilten:2016tkc}).

\subsection{Beam Dumps}

Approximate limits are set for beam-dump experiments using Eq.~\ref{eq:t0}, where the efficiencies are determined using Eq.~\eqref{eq:eff_dump}.

\subsubsection{Electron Beam Dumps}

Limits on dark photons have been set in Refs.~\cite{Bjorken:2009mm,Andreas:2012mt} using data from the E141, E137, E774, KEK, and Orsay electron beam-dump experiments~\cite{Riordan:1987aw,Bjorken:1988as,Bross:1989mp,Konaka:1986cb,Davier:1989wz}.
Recasting these for an $X$ boson requires solving
\begin{equation}
  	(g_X x_e)^2 \mathcal{B}(X \to e^+e^-) \epsilon[\tau_X(g_X)]
	\geq 	(\varepsilon_{\rm max} e)^2 \mathcal{B}(\aprime \to e^+e^-) \epsilon[\tau_{\aprime}(\varepsilon_{\rm max})]  \, ,
\end{equation}
at each mass.
{\em N.b.}, all of these experiments were only sensitive to decays into electrons and photons.
In addition, recently the NA64 collaboration published long-lived \aprime constraints using $\aprime\to e^+e^-$~\cite{Banerjee:2018vgk}.
The length of the shielding (provided by a calorimeter) changed during the run, but on average it was about 0.25\,m.
The total decay volume (before the electromagnetic calorimeter) was 3.5\,m; however, to satisfy the selection criteria, the decay needed to happen prior to the first tracking station, which was about 1\,m from the shielding during this run.

\subsubsection{Proton Beam Dumps}

Limits on $\aprime \to e^+e^-$ decays have been set by the following experiments: $\nu$-CAL~I~\cite{Blumlein:1990ay,Blumlein:1991xh}, using $\pi^0\to\aprime\gamma$ decays~\cite{Blumlein:2011mv} and proton bremsstrahlung~\cite{Blumlein:2013cua}; 
CHARM~\cite{Bergsma:1985qz}, using $\eta^{(\prime)}\to\aprime\gamma$ decays~\cite{Gninenko:2012eq};
and NOMAD~\cite{Astier:2001ck} and PS191~\cite{Bernardi:1985ny} using $\pi^0\to\aprime\gamma$ decays~\cite{Gninenko:2011uv}.
Recasting these for an $X$ boson involves solving
\begin{equation}
  	\Gamma_{P\to X\gamma}(g_X) \mathcal{B}(X \to e^+e^-) \epsilon[\tau_X(g_X)]
	\geq \Gamma_{P\to \aprime\gamma}(\varepsilon_{\rm max}) \mathcal{B}(\aprime \to e^+e^-) \epsilon[\tau_{\aprime}(\varepsilon_{\rm max})] \, ,
\end{equation}
where $P=\pi^0$, $\eta$, or $\eta^{\prime}$ for meson-decay production, and
\begin{equation}
  	g_X^2 (2x_u + x_d)^2 \mathcal{B}(X \to e^+e^-) \epsilon[\tau_X(g_X)]
	\geq (\varepsilon_{\rm max} e)^2 \mathcal{B}(\aprime \to e^+e^-) \epsilon[\tau_{\aprime}(\varepsilon_{\rm max})]  \, ,
\end{equation}
for proton bremsstrahlung.

\subsection{LEP}

Mono-photon searches from LEP~\cite{Abdallah:2003np,Abdallah:2008aa} were used to set limits on dark photons that decay invisibly in Ref.~\cite{Fox:2011fx}.
Here, we assume on-shell \aprime production, and rescale the results of Ref.~\cite{Fox:2011fx} assuming $g_{\chi} \gg g_e$ and $m_{\chi} \ll \ma$.
Since Ref.~\cite{Fox:2011fx} only reports results for $m_X=10, 50$ and 100\,GeV, we simply interpolate to obtain results for other masses.

\bibliographystyle{JHEP}
\bibliography{bib}

\end{document}